\documentclass[]{rmaa}
\usepackage{natbib}

\usepackage{paralist}
\usepackage[T1]{fontenc}

\usepackage{psfrag,color}

\usepackage[latin1]{inputenc}
\usepackage{epsfig}
\usepackage{amsmath}
\usepackage{float}
\usepackage{graphicx} 
\usepackage{subfig}
\usepackage{multirow}
\usepackage{array}
\usepackage{rotating}
\usepackage[normalem]{ulem}
\usepackage{wasysym}
\usepackage{caption}
\usepackage{subcaption}

\title{CCD PHOTOMETRY OF TRAPEZIA STARS I\altaffilmark{1}}

\author{A. Ruelas-Mayorga\altaffilmark{2}, L. J. S\'anchez\altaffilmark{2}, A. P\'aez-Amador \altaffilmark{2}, O. Segura-Montero \altaffilmark{2}, A. Nigoche-Netro\altaffilmark{3}}

\altaffiltext{1}{Based upon observations acquired at the Observatorio Astron\'omico Nacional on the Sierra San Pedro M\'artir (OAN-SPM), Baja California, M\'exico.}
\altaffiltext{2}{Instituto de
Astronom\'{\i}a, Universidad Nacional Aut\'onoma de M\'exico,
M\'exico D.F., M\'exico.}
\altaffiltext{3}{Instituto de Astronom{\'\i}a y Meteorolog{\'\i}a, Universidad
     de Guadalajara, Guadalajara, Jal. 44130, M\'exico.}

\fulladdresses{
 \item A. Ruelas-Mayorga, L.~J.~S\'anchez, A. Pa\'ez-Amador, O. Segura-Montero,: Instituto de Astronom\'{\i}a, Universidad
        Nacional Aut\'onoma de M\'exico, Apartado Postal 70-264, Cd.
        Universitaria 04510, M\'exico D.F., M\'exico.
        (\email{rarm,leonardo,emacias,osegura@astro.unam.mx}).
\item A. Nigoche-Netro: Instituto de Astronom{\'\i}a y Meteorolog{\'\i}a, Universidad
                                                                          de Guadalajara, Guadalajara, Jal. 44130, M\'exico. (\email{anigoche@gmail.com}).}

\shortauthor{Ruelas-Mayorga et al.}
\shorttitle{CCD Photometry of Trapezia Stars I}

\abstract{We present photometric CCD observations of stars in four stellar trapezia ADS 15184, ADS 4728, ADS 2843, and ADS 16795. This study is performed on images obtained at the Observatorio Astron\'omico Nacional at San Pedro M\'artir (OAN), Baja California, M\'exico. In this work we utilise aperture photometry to measure the $U$, $B$, $V$, $R$ and $I$ magnitudes of some of the stars in these dynamically unstable stellar clusters (trapezia). 

Using the $Q=(U-B)-0.72(B-V)$ parameter we obtained the spectral type of the studied stars as well as their distance to the Sun and their reddening. Slight differences between the $Q$-derived Spectral types and those listed in SIMBAD might be due to a different value, from $0.72$, for the slope of the reddening line on the two-colour diagram.}

\resumen{Presentamos fotometr\'ia CCD de las estrellas en cuatro trapecios estelares ADS 15184, ADS 728, ADS 2843, y ADS 16795. El estudio se hace a partir de im\'agenes tomadas en el Observatorio Astron\'omico Nacional de San Pedro M\'artir (OAN), Baja California, M\'exico. El presente trabajo utiliza la t\'ecnica de fotomet\'ia de apertura para medir las magnitudes $U$, $B$, $V$, $R$ e $I$ de algunas estrellas en estos c\'umulos abiertos din\'amicamente inestables (trapecios). 

Usando el par\'ametro  $Q=(U-B)-0.72(B-V)$ se obtuvo el tipo espectral de las estrellas estudiadas, tambi\'en se derivaron su distancia al Sol y su enrrojecimiento. Ligeras diferencias en los tipos espectrales derivados a partir del par\'ametro $Q$ con los listados en SIMBAD,  podr\'ian deberse a un valor diferente de $0.72$ de la pendiente de la l\'inea de enrrojecimiento en el diagrama de dos colores.}

\addkeyword{Galaxy: stellar content, open clusters and associations: individual (ADS15184, ADS728, ADS2843, ADS16795), techniques: photometry}

\begin{document}

\maketitle

%%%%%%%%%%%%%%%%%%%%%%%%%%%%%%%%%%%%%%%%%%%%%%%%%%%%%%%%%%%%%%%%%%%%%%%%%%%%%%%%%%%%%%%%%

\section{INTRODUCTION}

\label{sec:intro}

The stellar trapezia are formed in the interior of emission nebulae, such as the Orion Nebula. They are physical systems formed by three or more approximately equal stars, where the largest separation  between its stars is never larger than three times the smallest separation \citep{Ambartsumian1955}. That means the distances between the stars that form a trapezium are of the same order of magnitude. Trapezia are completely different to the hierarchical systems, where there could be a difference of an order of magnitude (10) between the smallest and larger separations.

\begin{figure}[!htbp]
       \centering
       \includegraphics[width=11cm]{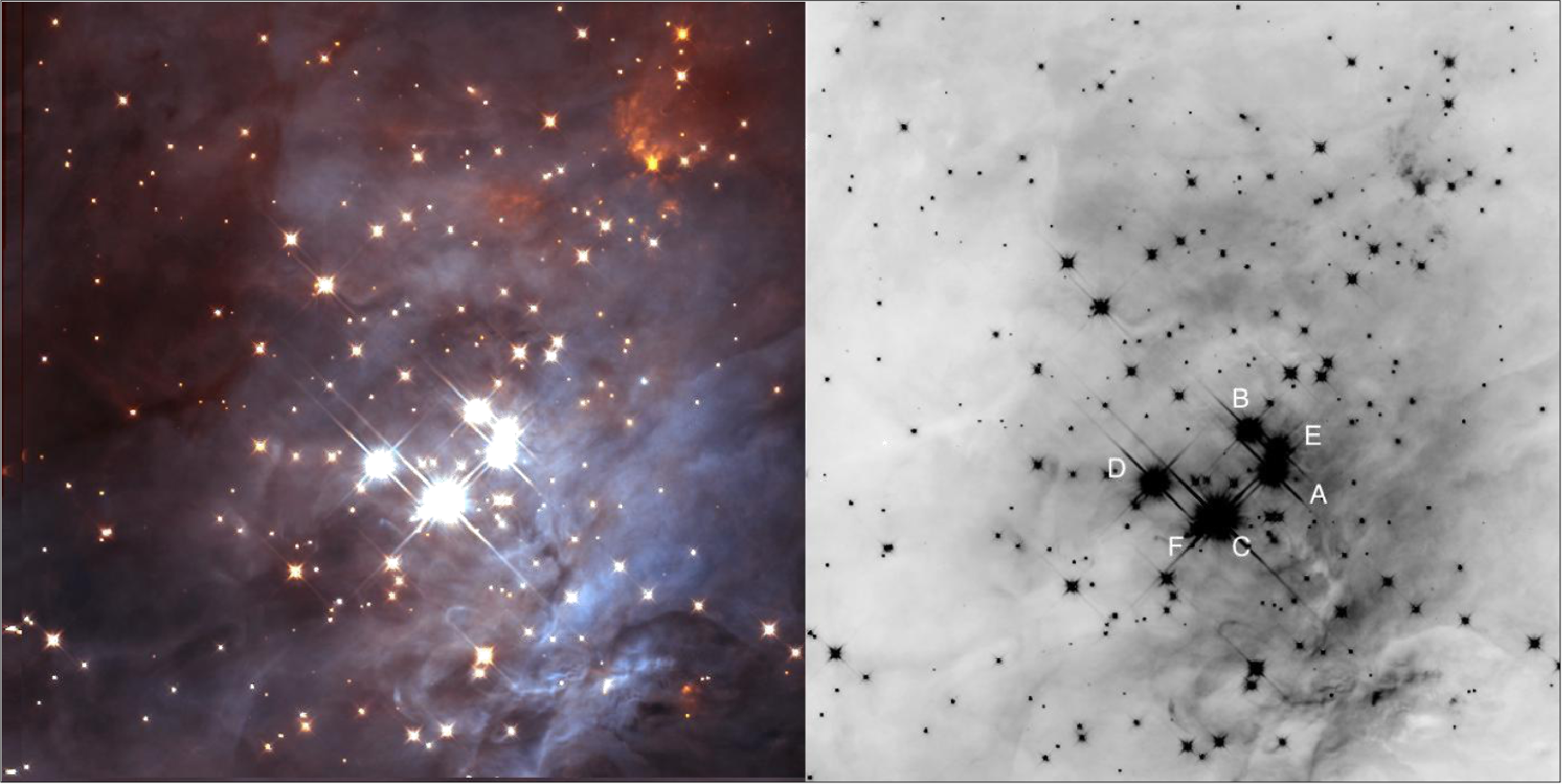}
       \caption{The prototypical trapezium type system in the Orion Nebula. On the right, the six main components are marked (Taken from \citet{Allen2019}).}
       \label{fig:TrapecioOrion}
   \end{figure}

The most well known trapezium system is found in the Orion Nebula. Figure \ref{fig:TrapecioOrion} shows this prototypical system. The brightest star in this system is $\theta^1$ C Orionis, around which we can see the other three stars in the system. These stars are very close at an approximate distance between them of  $\sim 1 \, AU$.

Trapezia are not dynamically stable, the orbits of their stellar components are not closed, this leads very quickly to close encounters which result in the expulsion of one or more members of the system and by this it turns into a hierarchical system \citep{Abt2000}. As a consequence of this fact it is found that the maximum age of stellar trapezia could not be larger than a few million years.

Using numerical simulation, \citet{Allen2018} showed that stellar trapezia evolve in time in different ways according to their initial configuration; some systems could break up into individual stars, whilst others evolve into binary or more complex stellar systems.

The age and evolution of trapezia also help us understand the evolution of stars.  Knowing the distance between their stars and the total size of a trapezium system is fundamental in order to understand its dynamical evolution \citep{Abt1986}.

%%%%%%%%%%%%%%%%%%%%%%%%%%%%%%%%%%%%%%%%%%%%%%%%%%%%%%%%%%%%%%%%%%%%%%%%%%%%%%%%%%%%%%%%%

%%%%%%%%%%%%%%%%%%%%%%%%%%%%%%%%%%%%%%%%%%%%%%%%%%%%%%%%%%%%%%%%%%%%%%%%%%%%%%%%%%%%%%%%%

This paper constitutes the first part of a photometric, spectroscopic and dynamical study of trapezia in our Galaxy. In Section \ref{sec:obs} we present the observations of the standard stars as well as those of the stars in the trapezia, in Section \ref{sec:slope} we discuss the slope of the reddening line on the two-colour diagram, and in Section \ref{sec:conclusions} we present our conclusions.

%%%%%%%%%%%%%%%%%%%%%%%%%%%%%%%%%%%%%%%%%%%%%%%%%%%%%%%%%%%%%%%%%%%%%%%%%%%%%%%%%%%%%%%%%

\section{The Observations}
\label{sec:obs}

The observations were taken during two observing seasons in June and December 2019, with the $84$ cm telescope at the OAN in San Pedro M\'artir, Baja California, M\'exico. % Tables \ref{tab:bitacorajune} and \ref{tab:bitacoradecember} show the way in which the observations  were performed.
We observed regions of standard stars \citep{Landolt1992} in the five filters of interest ($U$--$I$) distributed along the night as to have measurements at different values of Air Mass. Interspersed within these observations, we observed the stars in the trapezia of interest. The images have a plate scale of $0.47 \pm 0.06 \, arcsec/pix$ and a total size of $8.04 \pm 1.05 \, arcmin$ respectively. % as indicated in Tables \ref{tab:bitacorajune} and \ref{tab:bitacoradecember}.

%%%%%%%%%%%%%%%%%%%%%%%%%%%%%%%%%%%%%%%%%%%%%%%%%%%%%%%%%%%%%%%%%%%%%%%%%%%%%%%%%%%%%%%%%

\subsection{Reduction of standard stars}
\label{subsec:standardstars}

\begin{figure}[!htbp]
     \centering
     \includegraphics[width=10cm]{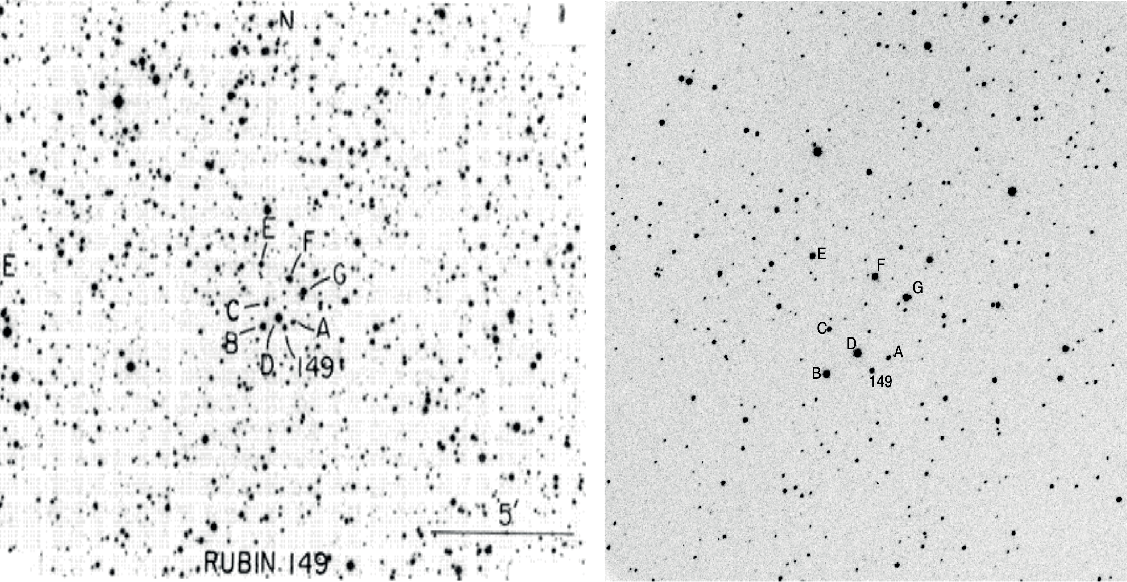}
     \caption{On the left the Standard region RUBIN 149 \citet{Landolt1992}, on the right our observed image.}
     \label{fig:RU149AB}
    \end{figure}

The Standard Stars are used in general to calibrate other astronomical observations. In this work, we shall refer to the set of equatorial standard stars published by \citet{Landolt1992}. We observed several of the Landolt standard regions every night and used them to calibrate our observations of the trapezia stars. Figure \ref{fig:RU149AB} shows a comparison of the standard region Rubin 149 observed by \citet{Landolt1992} and by us.

%%%%%%%%%%%%%%%%%%%%%%%%%%%%%%%%%%%%%%%%%%%%%%%%%%%%%%%%%%%%%%%%%%%%%%%%%%%%%%%%%%%%%%%%%

%%%%%%%%%%%%%%%%%%%%%%%%%%%%%%%%%%%%%%%%%%%%%%%%%%%%%%%%%%%%%%%%%%%%%%%%%%%%%%%%%%%%%%%%%

In order to express the magnitude of the stars in the trapezia in a standard system, we performed aperture photometry
of stars in some of the Landolt Standard Regions \citep{Landolt1992}. The photometric measurements of the standard stars was carried out using
the APT (Aperture Photometry Tool) programme, (see https://www.aperturephotometry.org/about/), and \citet{Laheretal2012}.

The APT programme is a software that permits aperture photometry measurements of stellar images on a frame. The programme accepts images in the \textit{fits} format, therefore no transformation of the images to other formats was necessary. The observed standard regions had already been pre-processed. 

The transformation equations between the observed and intrinsic photometric systems are as follows:

 \begin{equation}
    U_{int}-U_{obs}=A_{U}*X+K_U*(U-B)_{obs}+C_{U}
    \label{eq:MC0}
 \end{equation}
 \begin{equation}
    B_{int}-B_{obs}=A_{B}*X+K_B*(B-V)_{obs}+C_{B}
    \label{eq:MC1}
 \end{equation}
 \begin{equation}
    V_{int}-V_{obs}=A_{V}*X+K_V*(B-V)_{obs}+C_{V}
    \label{eq:MC2}
 \end{equation}
 \begin{equation}
    R_{int}-R_{obs}=A_{R}*X+K_R*(R-I)_{obs}+C_{R}
    \label{eq:MC3}
 \end{equation}
 \begin{equation}
    I_{int}-I_{obs}=A_{I}*X+K_I*(R-I)_{obs}+C_{I}
    \label{eq:MC4}
 \end{equation}

    where the suffixes \textit{int} and \textit{obs} stand for \textit{intrinsic} and \textit{observed}. The coefficients $A$, $K$ and $C$ rrepresent the negative of the coefficient of atmospheric absorption, a colour term and a zero point shift term. We intend to calculate these terms using the observed values measured with APT, and the intrinsic values for the magnitudes given in \citet{Landolt1992}. The Least Squares procedure is used to solve the equations, and the coefficients obtained are shown in Tables \ref{tab:Coeficientes U}--\ref{tab:Coeficientes I}.

%%%%%%%%%%%%%%%%%%%%%%%%%%%%%%%%%%%%%%%%%%%%%%%%%%%%%%%%%%%%%%%%%%%%%%%%%%%%%%%%%%%%%%%%%

    \begin{table}[!htbp]
    \caption{Transformation Coefficients for $U$}
    \begin{center}
    \begin{tabular}{  c  c  c  c  } \hline
    Standard Systems & A & K & C \\ \hline
     13-14 June      2019    &	-0.56	&	0.13	&	23.51	\\
     10-11  December 2019	&	-0.44	&	0.18	&	23.49	\\
     12-13  December 2019	&	-1.20	&	0.06	&	24.47	\\
     13-14  December 2019	&	-	    &	-	    &	-	    \\
     14-15  December 2019	&	17.07	&	-85.51	&	46.00	\\
     15-16  December 2019	&	-0.49	&	0.14	&	23.60	\\
    \hline
    \end{tabular}

    \label{tab:Coeficientes U}
    \end{center}
    \end{table}

%%%%%%%%%%%%%%%%%%%%%%%%%%%%%%%%%%%%%%%%%%%%%%%%%%%%%%%%%%%%%%%%%%%%%%%%%%%%%%%%%%%%%%%%%

    \begin{table}[!htbp]
    \caption{Transformation Coefficients for $B$}
    \begin{center}
    \begin{tabular}{  c  c  c  c  } \hline
    Standard Systems & A & K & C \\ \hline
    13-14  June     2019	&	-0.28	&	0.04	&	25.42	\\
    10-11  December 2019	&	-0.26	&	0.04	&	25.39	\\
    12-13  December 2019	&	-0.30	&	0.02	&	25.40	\\
    13-14  December 2019	&	664.84	&	3.79	&	-693.02	\\
    14-15  December 2019	&	-0.14	&	-0.08	&	25.14	\\
    15-16  December 2019	&	-0.31	&	0.04	&	25.42	\\
    \hline
    \end{tabular}
    \label{tab:Coeficientes B}
    \end{center}
    \end{table}

%%%%%%%%%%%%%%%%%%%%%%%%%%%%%%%%%%%%%%%%%%%%%%%%%%%%%%%%%%%%%%%%%%%%%%%%%%%%%%%%%%%%%%%%%

    \begin{table}[!htbp]
    \caption{Transformation Coefficients for $V$}
    \begin{center}
    \begin{tabular}{  c  c  c  c  } \hline
    Standard Systems & A & K & C \\ \hline
     13-14  June     2019    &	-0.17	&	-0.07	&	25.04	\\
     10-11  December 2019	&	-0.14	&	-0.07	&	25.05	\\
     12-13  December 2019	&	-0.10	&	-0.08	&	24.97	\\
     13-14  December 2019	&	-39.46	&	0.75	&	63.66	\\
     14-15  December 2019	&	-0.14	&	-0.09	&	25.04	\\
     15-16  December 2019	&	-0.17	&	-0.06	&	25.10	\\
    \hline
    \end{tabular}
    \label{tab:Coeficientes V}
    \end{center}
    \end{table}

%%%%%%%%%%%%%%%%%%%%%%%%%%%%%%%%%%%%%%%%%%%%%%%%%%%%%%%%%%%%%%%%%%%%%%%%%%%%%%%%%%%%%%%%%

    \begin{table}[!htbp]
    \caption{Transformation Coefficients for $R$}
    \begin{center}
    \begin{tabular}{  c  c  c  c  } \hline
    Standard Systems & A & K & C \\ \hline
     13-14  June     2019    &	-0.14	&	-0.05	&	25.09	\\
     10-11  December 2019	&	-0.10	&	-0.07	&	25.16	\\
     12-13  December 2019	&	-0.14	&	-0.03	&	25.15	\\
     13-14  December 2019	&	19.33	&	-2.21	&	5.28	\\
     14-15  December 2019	&	-0.09	&	-0.07	&	25.10	\\
     15-16  December 2019	&	-0.19	&	-0.07	&	25.22	\\
    \hline
    \end{tabular}
    \label{tab:Coeficientes R}
    \end{center}
    \end{table}

%%%%%%%%%%%%%%%%%%%%%%%%%%%%%%%%%%%%%%%%%%%%%%%%%%%%%%%%%%%%%%%%%%%%%%%%%%%%%%%%%%%%%%%%%

    \begin{table}[!htbp]
    \caption{Transformation Coefficients for $I$}
    \begin{center}
    \begin{tabular}{  c  c  c  c  } \hline
    Standard Systems & A & K & C \\ \hline
     13-14  June      2019   &	-0.10	&	0.11	&	24.95	\\
     10-11  December  2019	&	-0.04	&	0.11	&	25.06	\\
     12-13  December  2019	&	-0.34	&	0.14	&	25.47	\\
     13-14  December  2019	&	-3.51	&	-1.37	&	29.93	\\
     14-15  December  2019	&	-0.06	&	0.07	&	25.11	\\
     15-16  December  2019	&	-0.12	&	0.05	&	25.17	\\
    \hline
    \end{tabular}
    \label{tab:Coeficientes I}
    \end{center}
    \end{table}

%%%%%%%%%%%%%%%%%%%%%%%%%%%%%%%%%%%%%%%%%%%%%%%%%%%%%%%%%%%%%%%%%%%%%%%%%%%%%%%%%%%%%%%%%

\vskip2.0cm

\begin{figure}[!htbp]
     \centering
     \includegraphics[width=14cm,height=18cm]{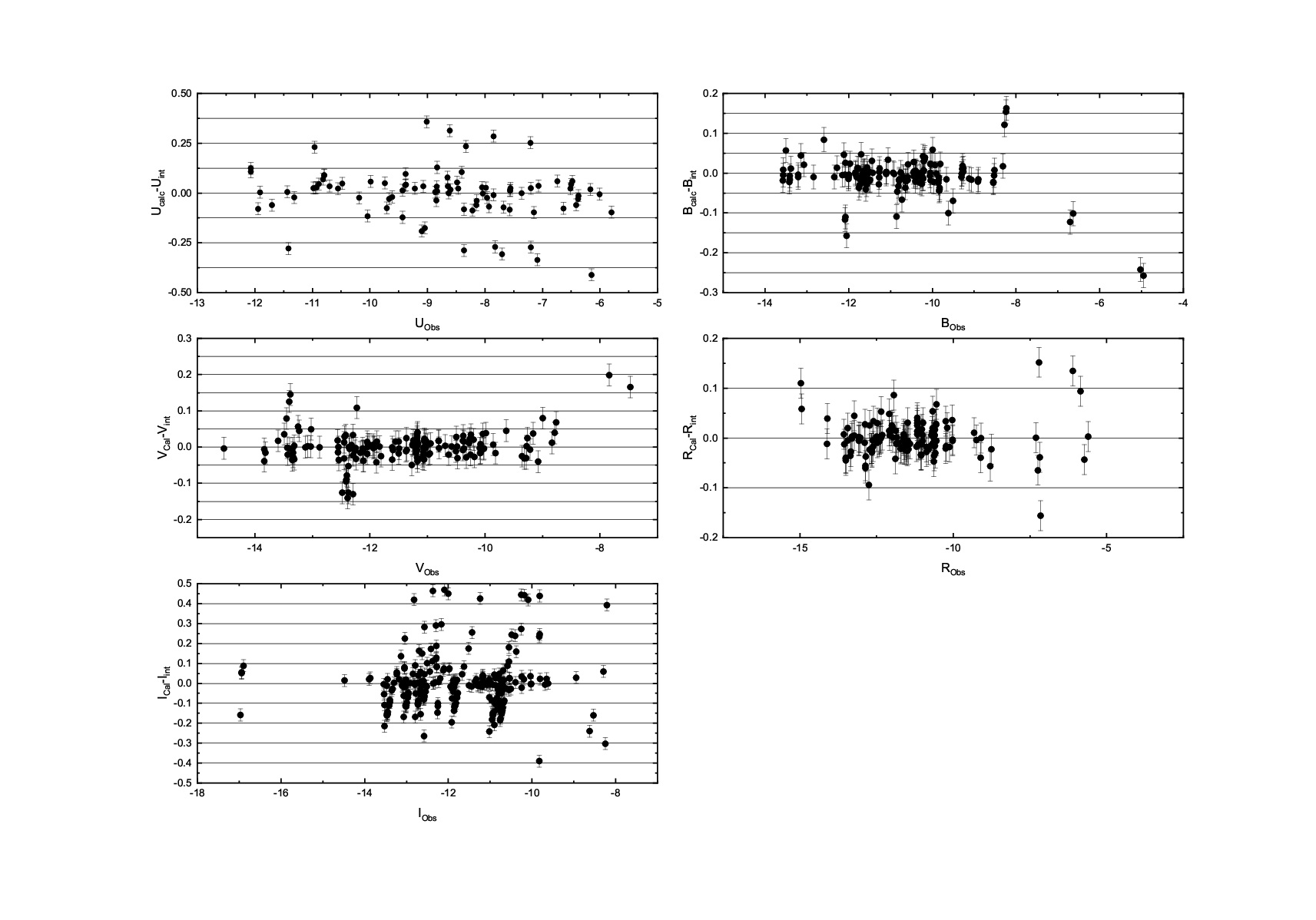}
     \caption{Calculated minus intrinsic magnitudes versus observed magnitudes for the $U$--$I$ filters.}
     \label{fig:MAGCALMAGINTvsMAGOBS}
\end{figure}

In Figure \ref{fig:MAGCALMAGINTvsMAGOBS}  we see graphs of the calculated magnitude minus the intrinsic magnitude versus the observed instrumental magnitude $(-2.5\,log(\# \, counts))$ for the five filters $U$--$I$, for the set of standard stars we use as calibrators. As can be seen from the Figures, the majority of the stars lie confined in the interval $-0.1 \leq M_{cal}-M_{int} \leq 0.1$, with a few stars falling outside this interval, except for the $I$ filter where a substantial number of stars ventures past the $\pm 0.2$ limit, reaching as far as $\geq 0.4$. When we check these values they all come from the observations on the nights $2019$, December $12$ and $13$ and from the Rubin $149$ and PG0220 standard regions. The night of December $12$ was an excellent and photometric observing night, whereas that on the $13$ of December was a very poor night, for which in fact, we have discarded the observations acquired. %In other investigations \citet{Ruelasetal2024} we have also encountered problems with standard region RU149.

%%%%%%%%%%%%%%%%%%%%%%%%%%%%%%%%%%%%%%%%%%%%%%%%%%%%%%%%%%%%%%%%%%%%%%%%%%%%%%%%%%%%%%%%%

%%%%%%%%%%%%%%%%%%%%%%%%%%%%%%%%%%%%%%%%%%%%%%%%%%%%%%%%%%%%%%%%%%%%%%%%%%%%%%%%%%%%%%%%%

\vfill

%%%%%%%%%%%%%%%%%%%%%%%%%%%%%%%%%%%%%%%%%%%%%%%%%%%%%%%%%%%%%%%%%%%%%%%%%%%%%%%%%%%%%%%%%

\subsection{Photometry of Trapezia Stars}
\label{subsec:trapeziastars}

Trapezia systems are open clusters in which their stars interact gravitationally, and this interaction has very noticeable dynamical consequences.  As mentioned in subsection  \ref{subsec:standardstars} the aperture photometry for the standard stars was done using the APT programme, however the photometry of the trapezia stars must be carried out on a set of more than a thousand images. For this photometric measurements, we used the aperture photometry capabilities of the programme  AstroImageJ (see http://astro.phy.vanderbilt.edu/~vida/aij.htm, \citet{AstroImageJCol}). As the aperture photometry measurements carried out with AstroImageJ are automatic, we performed a comparison between  the results obtained with APT and with AstroImageJ for the stars in one trapezium. Table   \ref{tab:AstroVsApt} shows a comparison  between the number of counts obtained with AstroImageJ  and that found with APT. This was done for measurements of the six components (A, B, C, D, E, and F) of the trapezium ADS 13292 --the full photometry for this trapezium will be reported elsewhere--. As stated in the last column of this Table, the difference in the number of counts measured with one programme or the other is at most of the order $0.1\%$.

%%%%%%%%%%%%%%%%%%%%%%%%%%%%%%%%%%%%%%%%%%%%%%%%%%%%%%%%%%%%%%%%%%%%%%%%%%%%%%%%%%%%%%%%%

\begin{table}[!htbp]
\caption{Measured counts for components A to F of ADS 13292, obtained with AstroImageJ and APT.}
\begin{center}
\begin{tabular}{  c  c  c  c  c  } \hline
 & Number of Counts &   Number of Counts & Difference & Difference(\%) \\
 &  (AstroImageJ)          &  (APT)    &    (counts)  &   (Percentage)    \\    \hline
A& 163657.5663&	163640&	    17.566316&	    0.01073358012\\
B& 1505427.035&	1503700&	1727.034662&	0.1147205824\\
C& 72591.87383&	72488.6&	103.273825& 	0.1422663716\\
D& 58946.5638&	    59003&  	56.436205& 	0.09574129748\\
E& 123743.0788&	123762& 	18.921178& 	0.015290696\\
F& 1170320.58&	    1169260&	1060.580108&	0.09062304176\\
 &  &   &   & \\
A& 40307.63323&	40359.5&	51.866766& 	0.1286772798\\
B& 94842.91829&	94752.7&	90.218291&  	0.0951239087\\
C&871463.4746&	870522& 	941.474566& 	0.1080337379\\
D& 80068.28323&	80192.9&	124.616769&   	0.1556381178\\
E& 150244.2993&	150189& 	55.299313&	0.0368062637\\
F& 1175990.939&	1175830&	160.93914&	0.01368540646\\ \hline
\end{tabular}
\label{tab:AstroVsApt}
\end{center}
\end{table}

%%%%%%%%%%%%%%%%%%%%%%%%%%%%%%%%%%%%%%%%%%%%%%%%%%%%%%%%%%%%%%%%%%%%%%%%%%%%%%%%%%%%%%%%%

We opted for using the AstroImageJ programme for the measurement of the stellar components of the different trapezia since the measurement of a large number of images may be done in an automatic way.

\vfill

%%%%%%%%%%%%%%%%%%%%%%%%%%%%%%%%%%%%%%%%%%%%%%%%%%%%%%%%%%%%%%%%%%%%%%%%%%%%%%%%%%%%%%%%%

\subsubsection{ADS 15184}
\label{subsubsec:ads15184}

\begin{figure}[!htbp]
     \centering
     \includegraphics[width=13cm]{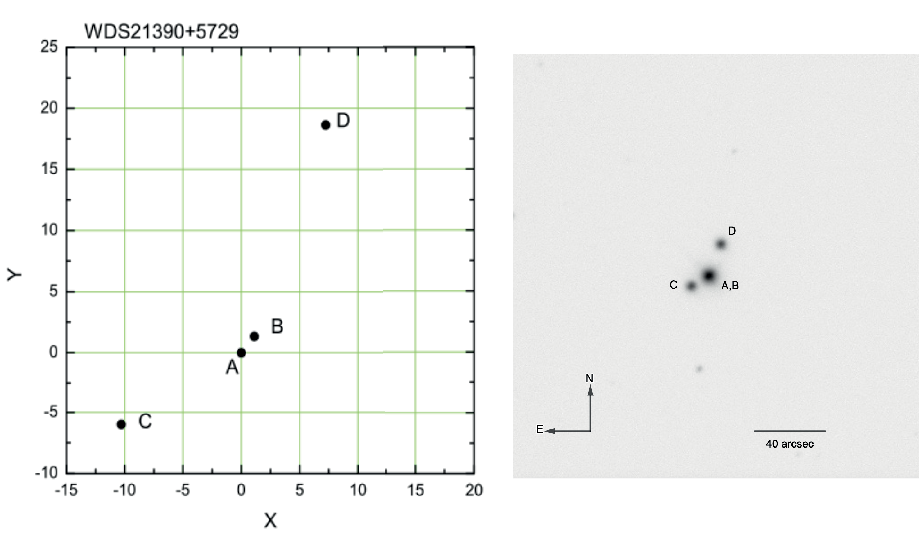}
     \caption{The left image represents the components of the trapezium ADS 15184. The axes are given in arcsecs. On the right, we show our observed image in the $V$ filter, with N-E orientation and scale in arsecs. Exposure time is 0.2 seconds}
     \label{fig:ComponentesTrapecioADS15184}
    \end{figure}

\begin{figure}[!htbp]
     \centering
     \includegraphics[width=13cm]{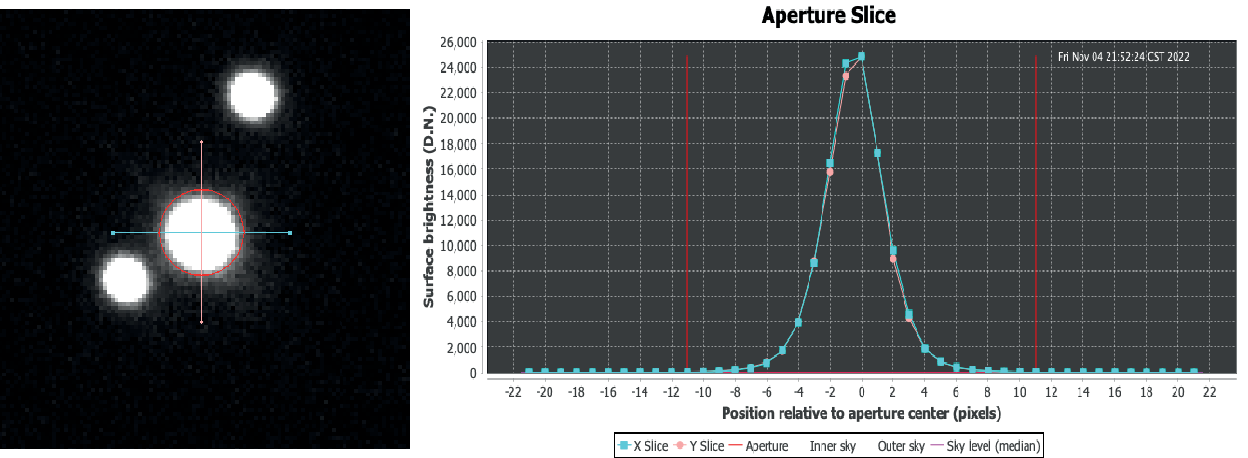}
     \caption{At the centre of the left image we show the blended $AB$ components of ADS 15184. On the right panel, vertical and horizontal cuts are performed on the central star component $AB$. These cuts show that components $A$ and $B$ are not well resolved in our images.}
     \label{fig:ComponentesCorteADS15184}
    \end{figure}

The trapezium ADS 15184 (also known as WDS 21390+5729) is located at RA=$21h$ $38m$, Dec=$+57^o$ $29\arcmin$. It has four bright components denoted as A, B, C, and D. Their distribution is presented on the left of Figure \ref{fig:ComponentesTrapecioADS15184}. The right panel of this Figure shows our image for ADS 15184 in the $V$ filter. On the figure we have identified the components $A$, $B$, $C$ and $D$. The $A$ and $B$ components are very bright and even on the short time exposure appear melded as shown in Figure \ref{fig:ComponentesCorteADS15184}. Due to this fact, the photometry of both these objects is carried out as if they represented one star only, which we denote as the $AB$ component.

We obtained 70 images of this trapezium in the filters $U-I$. Two sets of images were taken with integration times of $0.1$ and $8$ seconds respectively, so that we could get images where the bright stars were not saturated and also images where the fainter stars could have a substantial number of counts.

%%%%%%%%%%%%%%%%%%%%%%%%%%%%%%%%%%%%%%%%%%%%%%%%%%%%%%%%%%%%%%%%%%%%%%%%%%%%%%%%%%%%%%%%%

%%%%%%%%%%%%%%%%%%%%%%%%%%%%%%%%%%%%%%%%%%%%%%%%%%%%%%%%%%%%%%%%%%%%%%%%%%%%%%%%%%%%%%%%%

%%%%%%%%%%%%%%%%%%%%%%%%%%%%%%%%%%%%%%%%%%%%%%%%%%%%%%%%%%%%%%%%%%%%%%%%%%%%%%%%%%%%%%%%%

To perform photometry on each one of the stars of a trapezium, it is necessary to vary the aperture radius to ensure that the same percentage of the light is included for each stellar image. Table \ref{tab:RadiosADS15184} shows the radii for each component, where $r_{1}$ is the radius of the central aperture, $r_{2}$ is the inner radius of the sky annulus, and $r_{3}$ is the outer radius of the sky annulus.

%%%%%%%%%%%%%%%%%%%%%%%%%%%%%%%%%%%%%%%%%%%%%%%%%%%%%%%%%%%%%%%%%%%%%%%%%%%%%%%%%%%%%%%%%

    \begin{table}[!htbp]
     \caption{Photometric aperture radii in pixels for the trapezium ADS 15184.}
    \begin{center}
    \begin{tabular}{ c  c  c  c } \hline
     & $r_{1}$  & $r_{2}$   & $r_{3}$ \\
    \hline
      & $(pix)$  & $(pix)$   & $(pix)$ \\
    \hline
    AB& 8   & 12    & 16 \\
    C & 8   & 11    & 14 \\
    D & 8   & 11    & 14 \\
    \hline
    \end{tabular}
    \label{tab:RadiosADS15184}
    \end{center}
    \end{table}

%%%%%%%%%%%%%%%%%%%%%%%%%%%%%%%%%%%%%%%%%%%%%%%%%%%%%%%%%%%%%%%%%%%%%%%%%%%%%%%%%%%%%%%%%

The photometric measurements we obtained are given in number of counts, which is directly proportional to the exposure time. We normalise all our measurements to a standard time of $10$ seconds. To achieve this, we use the following equation:

    \begin{equation}
        I_{10}=\frac{10}{t}I_{obs}.
       \label{Ec:Inten10sec}
    \end{equation}

 Where $I_{10}$ represents the intensity of the star at the normalised time $(10 \, seconds)$, $t$ is the observation time and $I_{obs}$ is the observed intensity. The normalised intensity is transformed to magnitude in a standard manner.

Using equations \ref{eq:MC0}--\ref{eq:MC4} for each filter and taking the coefficients $A$, $K$ and $C$ for the night $10-11$ December, shown in Tables \ref{tab:Coeficientes U}--\ref{tab:Coeficientes I}, we calculate the values for the magnitudes and colours of the stars in the trapezium. We estimate the magnitude errors to be of the order of $\pm 0.02$ and $\pm 0.03$ for the colours.

%%%%%%%%%%%%%%%%%%%%%%%%%%%%%%%%%%%%%%%%%%%%%%%%%%%%%%%%%%%%%%%%%%%%%%%%%%%%%%%%%%%%%%%%%

    \begin{table}[!htbp]
    \caption{Observed Magnitudes and Colours for the stars in trapezium ADS15184.}
    \begin{center}
    \begin{tabular}{ c  c  c  c  c   c  c  c  c  c  c  } \hline
     &     U &     B &     V &     R &     I &   U-B &   B-V &   V-R &   R-I &   V-I \\
    \hline
    AB&  5.29 &  5.62 &  5.37 &  5.28 &  5.10 & -0.33 &  0.24 &  0.09 &  0.18 &  0.27 \\
    C &  7.93 &  8.18 &  7.87 &  7.66 &  7.49 & -0.25 &  0.32 &  0.20 &  0.17 &  0.37 \\
    D &  7.83 &  8.05 &  7.80 &  7.67 &  7.52 & -0.22 &  0.25 &  0.13 &  0.15 &  0.28 \\
    \hline
    \end{tabular}
    \label{tab:ColyMagObsADS15184}
    \end{center}
    \end{table}

%%%%%%%%%%%%%%%%%%%%%%%%%%%%%%%%%%%%%%%%%%%%%%%%%%%%%%%%%%%%%%%%%%%%%%%%%%%%%%%%%%%%%%%%%

Using the colour values, we may determine the $Q$ index (see Johnson \& Morgan (1953)), which we know is reddening independent ($Q=(U-B)-0.72(B-V)$).

We can associate a different spectral type to different values of $Q$ depending on whether we consider stars of Luminosity Class $I$ (Supergiants) or $V$ (Dwarfs) (see Table \ref{tab:Q contra ST}). We could also use for this classification the empirical calibration published by \citet{Lyubimkov2002} in which they associate the effective temperature of stars of luminosity classes II-III and IV-V to the value of the $Q$-parameter (see their eqs. (6) and (7) and their figure (11)). We shall try this approach elsewhere.

%%%%%%%%%%%%%%%%%%%%%%%%%%%%%%%%%%%%%%%%%%%%%%%%%%%%%%%%%%%%%%%%%%%%%%%%%%%%%%%%%%%%%%%%%

\begin{table}[!htbp]
\caption{Q versus Spectral Type. Taken from \citet{Johnson1953}.}
\begin{center}
\begin{tabular}{  c  c  c c} \hline
Spectral Type	&	Q	&	Spectral Type	&	Q	\\ \hline
O5	&	-0.93	&	B3	&	-0.57	\\
O6	&	-0.93	&	B5	&	-0.44	\\
O8	&	-0.93	&	B6	&	-0.37	\\
O9	&	-0.9	&	B7	&	-0.32	\\
B0	&	-0.9	&	B8	&	-0.27	\\
B0.5	&	-0.85	&	B9	&	-0.13	\\
B1	&	-0.78	&	A0	&	0	\\
B2	&	-0.7	&		&		\\
\hline
\end{tabular}
\label{tab:Q contra ST}
\end{center}
\end{table}

%%%%%%%%%%%%%%%%%%%%%%%%%%%%%%%%%%%%%%%%%%%%%%%%%%%%%%%%%%%%%%%%%%%%%%%%%%%%%%%%%%%%%%%%%

Table \ref{tab:SpectralTypeADS15184} gives the spectral types associated to each star of this trapezium based on the values of $Q$. In the case of $AB$, the spectral type given is that of a star that results from the union of A and B.

%%%%%%%%%%%%%%%%%%%%%%%%%%%%%%%%%%%%%%%%%%%%%%%%%%%%%%%%%%%%%%%%%%%%%%%%%%%%%%%%%%%%%%%%%

    \begin{table}[!htbp]
     \caption{Spectral types for Stars in ADS 15184.}
    \begin{center}
    \begin{tabular}{  c  c  c  } \hline
     &         Q & Spectral Type \\ \hline
    AB& -0.50 &             B5 \\
    C & -0.48 &             B5 \\
    D & -0.40 &             B6 \\
    \hline
    \end{tabular}

    \label{tab:SpectralTypeADS15184}
    \end{center}
    \end{table}

%%%%%%%%%%%%%%%%%%%%%%%%%%%%%%%%%%%%%%%%%%%%%%%%%%%%%%%%%%%%%%%%%%%%%%%%%%%%%%%%%%%%%%%%%

\begin{figure}[!htbp]
    \centering
    \includegraphics[width=10cm]{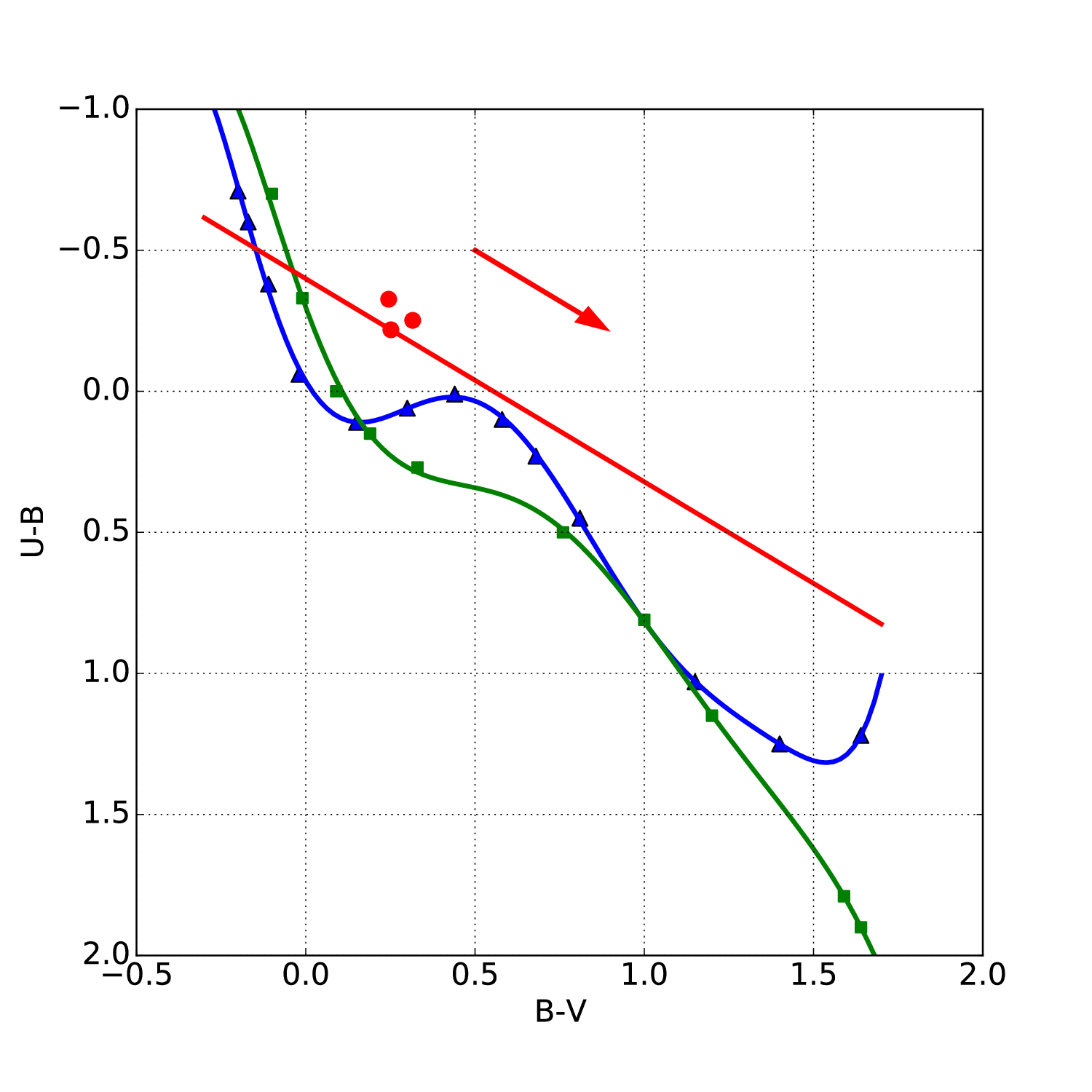}
    \caption{Two colour diagram $(U-B\, vs\, B-V)$ for ADS 15184. The green line is the Supergiant intrinsic sequence (SGIS), in blue the Main Sequence (MSIS) intrinsic sequence and the red dots represent the observed colours of the components of this trapezium. The red line and the red arrow represent the reddening line and direction.}
    \label{fig:colorcolorADS15184}
    \end{figure}

From the $U-B$ and $B-V$ colours shown in Table \ref{tab:ColyMagObsADS15184} we make a two-colour diagram, where the red dots represent the components $AB$, $C$ and $D$. The blue and green lines represent the intrinsic Main Sequence and the intrinsic Supergiant Sequence respectively. We also see the reddening line (red colour line, see Fig. \ref{fig:colorcolorADS15184}). This line has a slope equal to $\sim 0.72$ (see Fig \ref{fig:colorcolorADS15184}). This line allows us to deredden the observed colours of a star shifting the points in a direction parallel to this line until we intersect one of the intrinsic sequences (green or blue lines).

%%%%%%%%%%%%%%%%%%%%%%%%%%%%%%%%%%%%%%%%%%%%%%%%%%%%%%%%%%%%%%%%%%%%%%%%%%%%%%%%%%%%%%%%%

%%%%%%%%%%%%%%%%%%%%%%%%%%%%%%%%%%%%%%%%%%%%%%%%%%%%%%%%%%%%%%%%%%%%%%%%%%%%%%%%%%%%%%%%%

Finding the intersection of a line parallel to the reddening line that also passes on the observed points with the intrinsic sequences allows us to find the intrinsic colours of each star, which permits us to calculate the colour excess $E(B-V)$, the absorption $A_V$ and the distance in parsecs to the star in question ($d(pc)=10^{(m_{v}-M_{V}-A_{v}+5)/5}$). We have assumed the ratio of total to selective absorption $\left[\frac{A_V}{E(B-V)}\right]$ to be equal to $3.1$.

Table \ref{tab:DistanciaADS15184} gives the values for the intrinsic colours,  the excesses, the absorption and distances assuming the stars are supergiants (top panel), while in the bottom panel it gives the same information assuming the stars belong to the Main Sequence.

%%%%%%%%%%%%%%%%%%%%%%%%%%%%%%%%%%%%%%%%%%%%%%%%%%%%%%%%%%%%%%%%%%%%%%%%%%%%%%%%%%%%%%%%%

     \begin{table}[!htbp]
    \caption{Intrinsic colours, colour excess, absorption and distance for ADS 15184.}
    \begin{center}
%    \hline

    \begin{tabular}{  c  c  c  c  c  c  c  c} \hline
        \multicolumn{8}{c}{Assuming Supergiant Stars}\\
        \hline
   &       $(U-B)_{0}$&    $(B-V)_{0}$&    $E_{U-B}$&  $E_{B-V}$&  $A_{V}$&    Distance (pc)& Parallax (mas)\\
    \hline
    AB&   -0.56 &   -0.07 &    0.23 &    0.32 &  0.99 &   1380.43 &      0.72 \\
    C &   -0.53 &   -0.07 &    0.27 &    0.38 &  1.18 &   3975.93 &      0.25 \\
    D &   -0.43 &   -0.04 &    0.21 &    0.29 &  0.90 &   4349.30 &      0.23 \\
    \hline
       \multicolumn{8}{c}{Assuming Main Sequence Stars}\\
       \hline
   &       $(U-B)_{0}$&    $(B-V)_{0}$&    $E_{U-B}$&  $E_{B-V}$&  $A_{V}$&    Distance (pc)& Parallax (mas)\\
    \hline
    AB&   -0.63 &   -0.18 &    0.31 &    0.42 &  1.31 &    135.51 &      7.38 \\
    C &   -0.60 &   -0.17 &    0.35 &    0.49 &  1.51 &    389.58 &      2.57 \\
    D &   -0.51 &   -0.15 &    0.29 &    0.40 &  1.24 &    355.85 &      2.81 \\
    \hline
    \end{tabular}
    \label{tab:DistanciaADS15184}
    \end{center}
    \end{table}

%%%%%%%%%%%%%%%%%%%%%%%%%%%%%%%%%%%%%%%%%%%%%%%%%%%%%%%%%%%%%%%%%%%%%%%%%%%%%%%%%%%%%%%%%

Comparison of our results with the SIMBAD Astronomical Database \citep{SIMBAD2000} is presented in Table \ref{tab:ComparacionADS15184}. The magnitudes of the stars listed in SIMBAD come from the following references: \citet{Fabricius2002}, \citet{Reed2003}, \citet{Mercer2009}, and \citet{Zacharias2012}. Table \ref{tab:ComparacionADS15184} shows that our magnitudes differ in the worst case in one magnitude.

%%%%%%%%%%%%%%%%%%%%%%%%%%%%%%%%%%%%%%%%%%%%%%%%%%%%%%%%%%%%%%%%%%%%%%%%%%%%%%%%%%%%%%%%%

    \begin{table}[!htbp]
    \caption{Comparison of our measurements of ADS 15184 with SIMBAD (*).}
    \begin{center}
    \begin{tabular}{  c  c  c  c } \hline
    \multicolumn{4}{ c }{Magnitude} \\ \hline
     & AB   & C     & D     \\ \hline
    U & 5.29    & 7.93  & 7.83 \\
    U*& -       & 7.72  & 7.53 \\
    B & 5.62    & 8.18  & 8.05 \\
    B*& -       & 7.63  & -    \\
    V & 5.37    & 7.87  & 7.80 \\
    V*& -       & 7.46  & -    \\
    R & 5.28    & 7.66  & 7.67 \\
    R*& -       & -     & 8.73 \\
    I & 5.10    & 7.49  & 7.52 \\
    I*& -       & 7.67  & -    \\ \hline
    \multicolumn{4}{ c }{Spectral Type} \\ \hline
     & AB   & C     & D     \\ \hline
    St& B5      & B5    & B6   \\
    St*& -      & B1.5V & B1V  \\ \hline
    \multicolumn{4}{ c }{Parallax} \\ \hline
     & AB   & C     & D     \\ \hline
    $P_{sg}$ & 0.72  & 0.25  & 0.23 \\
    $P_{sp}$  & 7.38  & 2.57  & 2.81 \\
    P*          & -     & 0.8589& 1.1051\\ \hline
    \end{tabular}
    \label{tab:ComparacionADS15184}
    \end{center}
    \end{table}

%%%%%%%%%%%%%%%%%%%%%%%%%%%%%%%%%%%%%%%%%%%%%%%%%%%%%%%%%%%%%%%%%%%%%%%%%%%%%%%%%%%%%%%%%

Using Curve 15 of van de Hulst's  \citep{Johnson1968} we can obtain the dereddened values of the magnitudes for each component. The results of this exercise is shown in Table \ref{tab:ColyMagDesADS15184}, these values coincide with those reported in Table \ref{tab:DistanciaADS15184}.

%%%%%%%%%%%%%%%%%%%%%%%%%%%%%%%%%%%%%%%%%%%%%%%%%%%%%%%%%%%%%%%%%%%%%%%%%%%%%%%%%%%%%%%%%

    \begin{table}[!htbp]
    \caption{Dereddened magnitudes and colours for stars in ADS 15184 using van de Hulst curve 15.}
    \begin{center}
    \begin{tabular}{ c  c  c  c  c   c  c  c  c  c  c  } \hline
     \hline
    \multicolumn{11}{c}{Assuming Supergiant Stars}\\
    \hline
     &     U &     B &     V &     R &     I &   U-B &   B-V &   V-R &   R-I &   V-I \\
    \hline
    AB& 3.76    & 4.31  & 4.39  & 4.55  & 4.63  & -0.55 & -0.07 & -0.16 & -0.08 &   -0.25\\					
    C &	6.10	& 6.62	& 6.68	& 6.78	& 6.93	& -0.52	& -0.07	& -0.10	& -0.14	&   -0.25\\					
    D &	6.44	& 6.86	& 6.90	& 7.00	& 7.09	& -0.42	& -0.04	& -0.10	& -0.09	&	-0.19\\					
    \hline

    \multicolumn{11}{c}{Assuming Main Sequence Stars}\\
    \hline
     &     U &     B &     V &     R &     I &   U-B &   B-V &   V-R &   R-I &   V-I \\
    \hline
    AB& 3.25    & 3.88  & 4.06  & 4.31  & 4.48  & -0.63 & -0.18 & -0.25 & -0.17 & -0.42	\\				
    C &	5.58    & 6.18  & 6.35  & 6.54	& 6.77	& -0.60	& -0.17	& -0.19	& -0.23	& -0.42	\\			
    D &	5.91	& 6.41	& 6.56	& 6.75	& 6.93	& -0.50	& -0.15	& -0.19	& -0.18	& -0.37	\\
    \hline
    \end{tabular}
    \label{tab:ColyMagDesADS15184}
    \end{center}
    \end{table}

    \vskip10cm

%%%%%%%%%%%%%%%%%%%%%%%%%%%%%%%%%%%%%%%%%%%%%%%%%%%%%%%%%%%%%%%%%%%%%%%%%%%%%%%%%%%%%%%%%

\subsubsection{ADS 4728}
\label{subsubsec:ads4728}

\begin{figure}[!htbp]
     \centering
     \includegraphics[width=13cm]{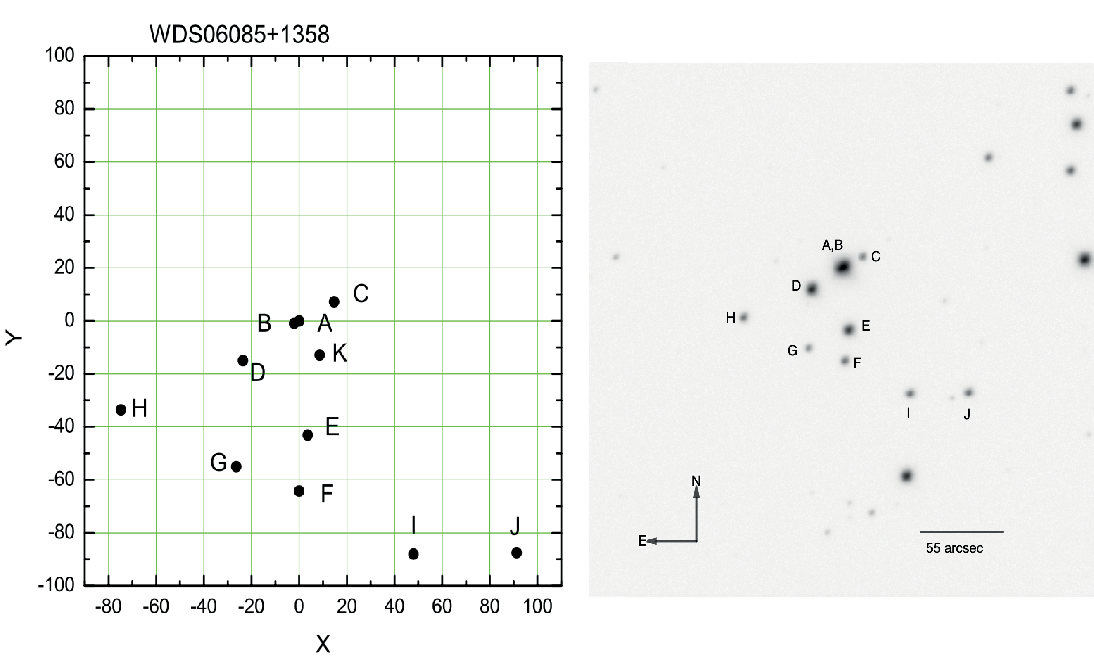}
     \caption{The left image represents the components of the trapezium ADS 4728. The axes units are seconds of arc. On the right, we show our observed image in the $V$ filter, with N-E orientation and scale in arsecs. Exposure time is 1 second.}
     \label{fig:ComponentsImageADS4728}
    \end{figure}

Trapezium ADS 4728 (also known as WDS 06085+13358) located at RA=$6h$ $8m$, Dec=$+13^o$ $58\arcmin$. It has eleven components: $A$--$K$, which are distributed as shown on the left panel of Figure \ref{fig:ComponentsImageADS4728}.

We observed 110 images for this trapezium in filters $U$--$I$.

The images for this trapezium were obtained with three different integration times $0.5$, $1$ and $8$ seconds.

Figure \ref{fig:ComponentsImageADS4728} also shows an image of this trapezium in the $V$ filter (right panel). The stellar components as well as the orientation and plate scale are indicated on this figure.

%Se puede observar de manera cualitativa que las componentes A y B no tienen la separación suficiente como para hacer fotometría de dichas estrellas por separado. Analizando con el programa APT (ver Figura \ref{fig:CorteADS4728}) y haciendo cortes verticales y horizontales similares a los presentados en la Figura \ref{fig:CorteADS15184} se puede verificar que además de no distinguir entre componentes, el CCD se encuentra saturado. Caso contrario para la componente K, la cual cuenta con una magnitud aparente baja, por lo cual el CCD no alcanza a recolectar los fotones necesarios para detectarla. Así, se decidió no hacer la fotometría de las componentes A, B y K.

%%%%%%%%%%%%%%%%%%%%%%%%%%%%%%%%%%%%%%%%%%%%%%%%%%%%%%%%%%%%%%%%%%%%%%%%%%%%%%%%%%%%%%%%%

Table \ref{tab:RadiosADS4728} shows the values of the radii we used for performing the photometry for each component. As done previously, all the measurements are normalised to $10$ seconds with Equation \ref{Ec:Inten10sec}.

%%%%%%%%%%%%%%%%%%%%%%%%%%%%%%%%%%%%%%%%%%%%%%%%%%%%%%%%%%%%%%%%%%%%%%%%%%%%%%%%%%%%%%%%%

    \begin{table}[!htbp]
    \caption{Photometric aperture radii for ADS 4728.}
    \begin{center}
    \begin{tabular}{ c  c  c  c } \hline
     & $r_{1}$  & $r_{2}$   & $r_{3}$ \\
     \hline
     & $(pix)$  & $(pix)$   & $(pix)$ \\
    \hline
    C & 9   & 11    & 13 \\
    D & 10  & 11    & 14 \\
    E & 9   & 11    & 14 \\
    F & 9   & 11    & 14 \\
    G & 11  & 13    & 16 \\
    H & 11  & 13    & 16 \\
    I & 10  & 12    & 14 \\
    J & 10  & 12    & 14 \\
    \hline
    \end{tabular}
    \label{tab:RadiosADS4728}
    \end{center}
    \end{table}

Using the values of $A$, $K$ and $C$ we obtain the magnitudes and colours for the stars of trapezium ADS 15184 (see Table \ref{tab:ColyMagObsADS4728}).
%para los cuales se estima un error de $\pm 0.02$ y $\pm 0.03$ respectivamente.

%%%%%%%%%%%%%%%%%%%%%%%%%%%%%%%%%%%%%%%%%%%%%%%%%%%%%%%%%%%%%%%%%%%%%%%%%%%%%%%%%%%%%%%%%

    \begin{table}[!htbp]
     \caption{Observed magnitudes and colours for trapezium  ADS 4728.}
    \begin{center}
    \begin{tabular}{ c  c  c  c  c  c  c  c  c  c  c  } \hline
     &     U &     B &     V &     R &     I &   U-B &   B-V &   V-R &   R-I &   V-I \\
    \hline
    C &  11.66 &  11.72 &  11.51 &  11.46 &  11.31 & -0.05 &  0.21 &  0.04 &  0.15 &  0.20 \\
    D &   7.98 &   8.60 &   8.59 &   8.64 &   8.62 & -0.62 &  0.01 & -0.05 &  0.02 & -0.03 \\
    E &   8.88 &   9.22 &   9.12 &   9.13 &   9.04 & -0.35 &  0.10 & -0.00 &  0.08 &  0.08 \\
    F &  11.16 &  11.08 &  10.87 &  10.83 &  10.68 &  0.08 &  0.21 &  0.04 &  0.15 &  0.19 \\
    G &  12.11 &  11.98 &  11.76 &  11.72 &  11.57 &  0.13 &  0.22 &  0.04 &  0.15 &  0.18 \\
    H &  12.25 &  11.08 &   9.88 &   9.34 &   8.76 &  1.17 &  1.21 &  0.54 &  0.58 &  1.12 \\
    I &  10.70 &  10.87 &  10.75 &  10.74 &  10.65 & -0.17 &  0.12 &  0.01 &  0.09 &  0.09 \\
    J &  10.72 &  10.95 &  10.81 &  10.76 &  10.64 & -0.22 &  0.13 &  0.05 &  0.12 &  0.17 \\
    \hline
    \end{tabular}
    \label{tab:ColyMagObsADS4728}
    \end{center}
    \end{table}

%%%%%%%%%%%%%%%%%%%%%%%%%%%%%%%%%%%%%%%%%%%%%%%%%%%%%%%%%%%%%%%%%%%%%%%%%%%%%%%%%%%%%%%%%

Table \ref{tab:SpectralTypeADS4728} shows the spectral type associated to each star from the $Q$ parameter calibration

%%%%%%%%%%%%%%%%%%%%%%%%%%%%%%%%%%%%%%%%%%%%%%%%%%%%%%%%%%%%%%%%%%%%%%%%%%%%%%%%%%%%%%%%%

    \begin{table}[!htbp]
    \caption{Q derived spectral types for ADS 4728}
    \begin{center}
    \begin{tabular}{  c  c  c  } \hline
     &         Q & Spectral Type \\ \hline
    C & -0.20 &             B8 \\
    D & -0.63 &             B3 \\
    E & -0.42 &             B5 \\
    F & -0.08 &             B9 \\
    G & -0.03 &             A0 \\
    H &  0.30 &             A0 \\
    I & -0.26 &             B8 \\
    J & -0.32 &             B7 \\
    \hline
    \end{tabular}
    \label{tab:SpectralTypeADS4728}
    \end{center}
    \end{table}

%%%%%%%%%%%%%%%%%%%%%%%%%%%%%%%%%%%%%%%%%%%%%%%%%%%%%%%%%%%%%%%%%%%%%%%%%%%%%%%%%%%%%%%%%

\begin{figure}[!htbp]
    \centering
    \includegraphics[width=10cm]{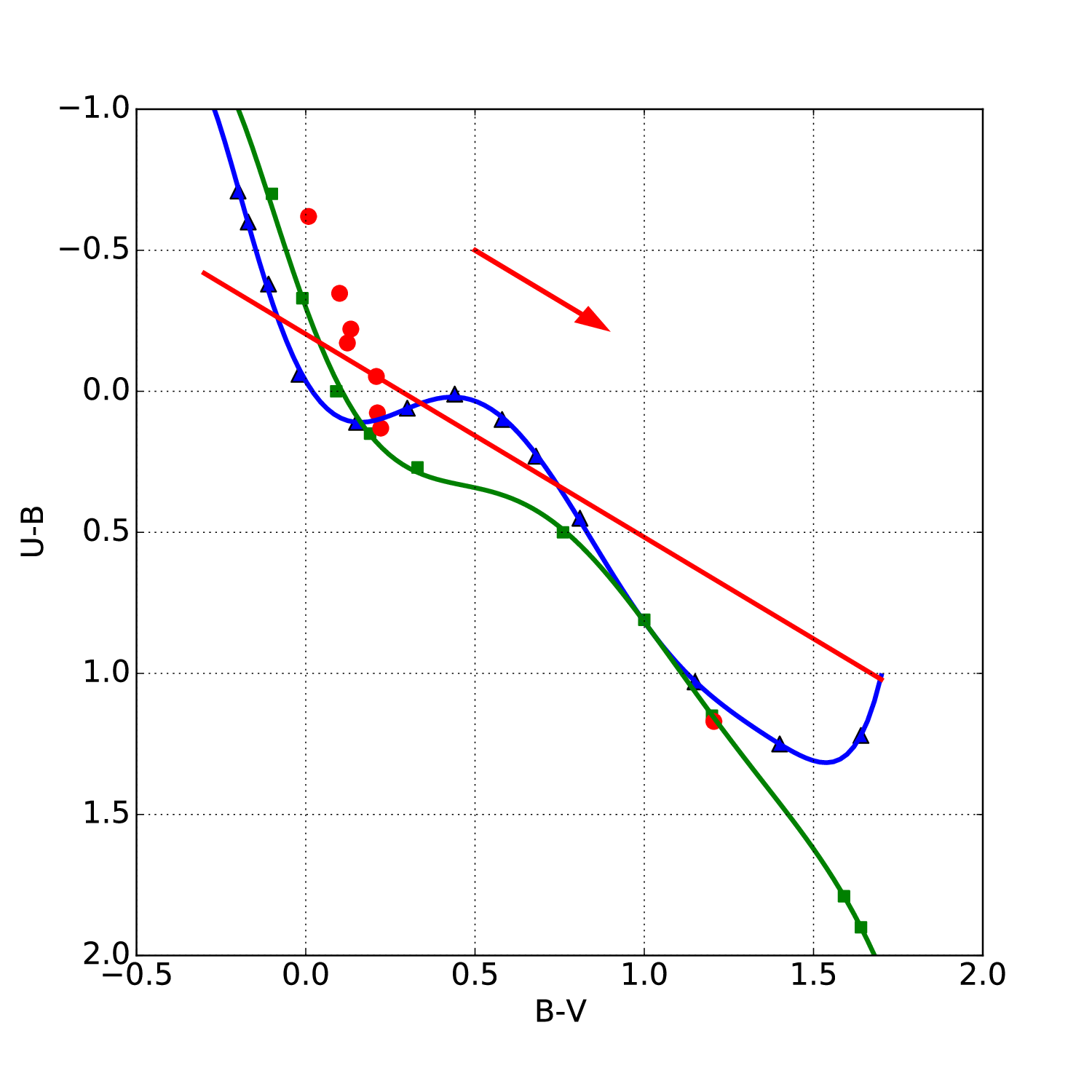}
    \caption{Two-colour diagram $(U-B \, vs \, B-V)$ for ADS 4728.}
    \label{fig:colorcolorADS4728}
    \end{figure}

Using the $U-B$ and $B-V$ colours from Table \ref{tab:ColyMagObsADS4728} we plot the points on the two-colour diagram (see Figure \ref{fig:colorcolorADS4728}) where the same conventions as those followed for the previous trapezium are followed.
%donde los puntos rojos representan las componentes C, D, E, F, G, H, I y J. En azul y verde se pueden encontrar las curvas de secuencia principal y de supergigantes respectivamente. Sobre el diagrama de dos colores se traza la l\'inea de enrojecimiento (l\'inea de color rojo en la Fig. \ref{fig:colorcolorADS4728}).

%%%%%%%%%%%%%%%%%%%%%%%%%%%%%%%%%%%%%%%%%%%%%%%%%%%%%%%%%%%%%%%%%%%%%%%%%%%%%%%%%%%%%%%%%

%%%%%%%%%%%%%%%%%%%%%%%%%%%%%%%%%%%%%%%%%%%%%%%%%%%%%%%%%%%%%%%%%%%%%%%%%%%%%%%%%%%%%%%%%

The first part of Table \ref{tab::DistanciasADS4728} presents the values for intrinsic colours, colour excesses, absorption and distances assuming the stars are supergiants, while the second part shows the same results assuming the stars to be in the Main Sequence.

%%%%%%%%%%%%%%%%%%%%%%%%%%%%%%%%%%%%%%%%%%%%%%%%%%%%%%%%%%%%%%%%%%%%%%%%%%%%%%%%%%%%%%%%%

    \begin{table}[!htbp]
     \caption{Intrinsic colours, colour excesses, Absorption and distances for ADS 4728.}
    \begin{center}
    \begin{tabular}{  c  c  c  c  c  c  c  c } \hline
    \hline
    \multicolumn{8}{c}{Assuming Supergiant Stars}\\
    \hline
      & $(U-B)_{0}$&    $(B-V)_{0}$&    $E_{B-V}$&  $E_{U-B}$&  $A_{V}$&    Distance (pc)& Parallax (mas)\\
    \hline
    C &   -0.17 &    0.04 &    0.17 &    0.12 &  0.52 &  27948.93 &      0.04 \\
    D &   -0.71 &   -0.11 &    0.12 &    0.09 &  0.38 &   8201.46 &      0.12 \\
    E &   -0.45 &   -0.04 &    0.14 &    0.10 &  0.45 &   9945.13 &      0.10 \\
    F &    0.00 &    0.11 &    0.10 &    0.07 &  0.32 &  22636.52 &      0.04 \\
    G &    0.07 &    0.14 &    0.08 &    0.06 &  0.25 &  34886.01 &      0.03 \\
    H &    - &    - &    - &    - &  - &   - &      - \\
    I &   -0.25 &    0.02 &    0.11 &    0.08 &  0.33 &  21454.01 &      0.05 \\
    J &   -0.32 &   -0.01 &    0.14 &    0.10 &  0.44 &  21340.38 &      0.05 \\
    \hline
    \multicolumn{8}{c}{Assuming Main Sequence Stars}\\
    \hline
      & $(U-B)_{0}$&    $(B-V)_{0}$&    $E_{B-V}$&  $E_{U-B}$&  $A_{V}$&    Distance (pc)& Parallax (mas)\\
      \hline
    C &   -0.26 &   -0.08 &    0.29 &    0.21 &  0.91 &   1466.26 &      0.68 \\
    D &   -0.78 &   -0.21 &    0.22 &    0.16 &  0.69 &   1261.05 &      0.79 \\
    E &   -0.53 &   -0.15 &    0.25 &    0.18 &  0.79 &    969.79 &      1.03 \\
    F &   -0.09 &   -0.03 &    0.24 &    0.17 &  0.74 &    931.28 &      1.07 \\
    G &   -0.02 &    0.01 &    0.22 &    0.16 &  0.67 &    866.56 &      1.15 \\
    H &   -     &   -     &    -    &    -    &  -    &    -      &      -    \\
    I &   -0.34 &   -0.10 &    0.23 &    0.16 &  0.71 &   1132.09 &      0.88 \\
    J &   -0.41 &   -0.12 &    0.26 &    0.18 &  0.80 &   1483.90 &      0.67 \\
    \hline
    \end{tabular}
    \label{tab::DistanciasADS4728}
    \end{center}
    \end{table}

%%%%%%%%%%%%%%%%%%%%%%%%%%%%%%%%%%%%%%%%%%%%%%%%%%%%%%%%%%%%%%%%%%%%%%%%%%%%%%%%%%%%%%%%%
The magnitudes of the stars listed in SIMBAD come from the following references: \citet{Hog2000}, \citet{Zacharias2003}, \citet{Reed2003}, \citet{Zacharias2009}, \citet{Krone-Martins2010}, and \citet{Gaia2020}.
Table \ref{tab:ComparacionADS4728} shows our results compared with those listed in SIMBAD (shown with *).

%%%%%%%%%%%%%%%%%%%%%%%%%%%%%%%%%%%%%%%%%%%%%%%%%%%%%%%%%%%%%%%%%%%%%%%%%%%%%%%%%%%%%%%%%

    \begin{table}[!htbp]
     \caption{Comparison of our results for ADS 4728 with SIMBAD (*).}
    \begin{center}
    \begin{tabular}{  c  c  c  c  c  c  c  c  c } \hline
    \multicolumn{9}{ c }{Magnitude} \\ \hline
      & C    & D     & E     & F     & G     & H     & I     & J \\ \hline
    U & 11.66 & 7.98 & 8.88 & 11.16 & 12.11 & 12.25 & 10.70 & 10.72\\
    U*& -     & 7.75 & -    & -     & 10.94 & -     & -     & -    \\
    B & 11.72 & 8.60 & 9.22 & 11.08 & 11.98 & 11.08 & 10.87 & 10.95\\
    B*& 11.51 & 8.239& 9.051& 11.02 & 11.882& 11.03 & 10.82 & -    \\
    V & 11.51 & 8.59 & 9.12 & 10.87 & 11.76 &  9.88 & 10.75 & 10.81\\
    V*& 11.529& 8.659& 9.265& 10.968& 11.892& 9.905 & 10.825& -    \\
    R & 11.46 & 8.64 & 9.13 & 10.83 & 11.72 &  9.34 & 10.74 & 10.76\\
    R*& 11.32 & 8.89 & 9.10 & 10.86 & 11.69 &  9.65 & -     & -    \\
    I & 11.31 & 8.63 & 9.04 & 10.68 & 11.57 &  8.76 & 10.65 & 10.64\\
    I*& -     & -    & -    & -     & 12.40 & -     & -     &  -   \\ \hline
    \multicolumn{9}{ c }{Spectral Type} \\ \hline
     & C    & D     & E     & F     & G     & H     & I     & J \\ \hline
    St& B8  & B3    & B5    & B9    & A0    & A0    & B8    & B7 \\
    St*&B9V & B1V   & B3V   & B9V   & -     & G8    & B5    & -  \\ \hline
    \multicolumn{9}{ c }{Parallax} \\ \hline
     & C    & D     & E     & F     & G     & H     & I     & J \\ \hline
    $P_{sg}$ & 0.04 & 0.12  & 0.10  & 0.04  & 0.03  & 0.24  & 0.05  & 0.05 \\
    $P_{sp}$ & 0.68 & 0.79  & 1.03  & 1.07  & 1.15  & 10.27 & 0.88  & 0.67 \\
    P*       & 1.0671 & 0.96 & 0.9537 & 1.0783 & 1.0151 & 1.657 & 1. 086 & -\\ \hline
    \end{tabular}
    \label{tab:ComparacionADS4728}
    \end{center}
    \end{table}

%%%%%%%%%%%%%%%%%%%%%%%%%%%%%%%%%%%%%%%%%%%%%%%%%%%%%%%%%%%%%%%%%%%%%%%%%%%%%%%%%%%%%%%%%

Table \ref{tab:ColyMagDesADS4728} shows the dereddened values for magnitudes and colours for the stars in ADS 4728.
%muestran los resultados de las magnitudes y colores desenrojecidos para estrellas supergigantes y de secuencia principal. En donde se observa que los datos de colores coinciden con los reportados en las Tablas \ref{tab:DistanciaADS4728} y \ref{tab:DistanciaADS4728}.

%%%%%%%%%%%%%%%%%%%%%%%%%%%%%%%%%%%%%%%%%%%%%%%%%%%%%%%%%%%%%%%%%%%%%%%%%%%%%%%%%%%%%%%%%

    \begin{table}[!htbp]
    \caption{Dereddened magnitudes and colours for ADS 4728 using van de Hulst curve 15.}
    \begin{center}
    \begin{tabular}{ c  c  c  c  c  c  c  c  c  c  c  }
    \hline
    \multicolumn{11}{c}{Assuming Supergiant Stars}\\
    \hline
     &     U &     B &     V &     R &     I &   U-B &   B-V &   V-R &   R-I &   V-I \\
    \hline
    C &	10.85   & 11.03	& 10.99	& 11.08	& 11.06	& -0.17	& 0.04	& -0.09	& 0.01	& -0.08 \\
    D &	7.39	& 8.1	& 8.21	& 8.36	& 8.44	& -0.71	& -0.11	& -0.15	& -0.08	& -0.23 \\
    E &	8.18	& 8.63	& 8.67	& 8.79	& 8.83	& -0.45	& -0.04	& -0.12	& -0.04	& -0.15 \\
    F &	10.66	& 10.66	& 10.55	& 10.59	& 10.53	& 0	    & 0.11	& -0.04	& 0.07	& 0.02  \\
    G &	11.73	& 11.65	& 11.51	& 11.54	& 11.46	& 0.07	& 0.14	& -0.03	& 0.08	& 0.06  \\
    H &	7.59	& 7.11	& 6.88	& 7.11	& 7.33	& 0.48	& 0.24	& -0.23	& -0.22	& -0.45 \\
    I &	10.18	& 10.43	& 10.41	& 10.49	& 10.49	& -0.25	& 0.02	& -0.08	& 0   	& -0.08 \\
    J&	10.05	& 10.37	& 10.38	& 10.44	& 10.43	& -0.32	& -0.01	& -0.06	& 0.01	& -0.05 \\
    \hline
    \multicolumn{11}{c}{Assuming Main Sequence Stars}\\
    \hline
     &     U &     B &     V &     R &     I &   U-B &   B-V &   V-R &   R-I &   V-I \\
    \hline
    C &	10.26	&	10.52	&	10.6	&	10.79	&	10.88	&	-0.26	&	-0.08	&	-0.19	&	-0.09	&	-0.28\\
    D &	6.91	&	7.69	&	7.9 	&	8.13	&	8.29	&	-0.78	&	-0.21	&	-0.23	&	-0.16	&	-0.39\\
    E &	7.65	&	8.18	&	8.33	&	8.54	&	8.67	&	-0.53	&	-0.15	&	-0.21	&	-0.13	&	-0.33\\
    F &	10.02	&	10.11	&	10.14	&	10.28	&	10.33	&	-0.09	&	-0.03	&	-0.15	&	-0.04	&	-0.19\\
    G &	11.07	&	11.09	&	11.09	&	11.22	&	11.25	&	-0.02	&	0.01	&	-0.13	&	-0.03	&	-0.17\\
    H &	6.77	&	6.41	&	6.34	&	6.71	&	7.07	&	0.36	&	0.06	&	-0.37	&	-0.36	&	-0.73\\
    I &	9.6	    &	9.93	&	10.04	&	10.22	&	10.31	&	-0.33	&	-0.1	&	-0.18	&	-0.1	&	-0.28\\
    J &	9.49	&	9.89	&	10.02	&	10.17	&	10.26	&	-0.4	&	-0.12	&	-0.15	&	-0.09	&	-0.24\\
    \hline
    \end{tabular}
    \label{tab:ColyMagDesADS4728}
    \end{center}
    \end{table}

\vskip10cm
\vfill

%%%%%%%%%%%%%%%%%%%%%%%%%%%%%%%%%%%%%%%%%%%%%%%%%%%%%%%%%%%%%%%%%%%%%%%%%%%%%%%%%%%%%%%%%

\subsubsection{ADS 2843}
\label{subsubsec:ads2843}

\begin{figure}[H]
    \centering
    \includegraphics[width=13cm]{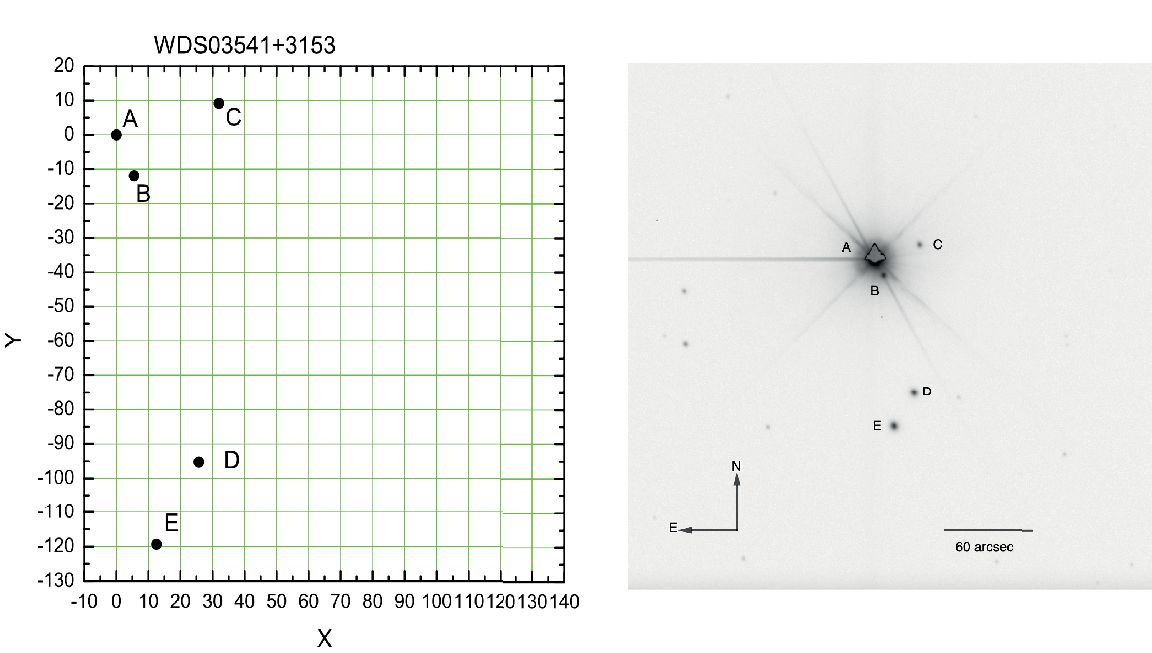}
    \caption{The left image represents the components of the trapezium ADS 2843. The axes units are in arcseconds. On the right, we show our observed image in the $V$ filter, with N-E orientation and scale in arsecs. $3$ seconds exposure time.}
    \label{fig:ComponentsImageADS2843}
    \end{figure}

Trapezium ADS 2843 (Also known as WDS 03541+3153) is at RA=$3h$ $54m$, Dec=$+31^o$ $53\arcmin$ and has five components: $A$, $B$, $C$, $D$, and $E$, distributed as seen on the left of Figure \ref{fig:ComponentsImageADS2843}.

%%%%%%%%%%%%%%%%%%%%%%%%%%%%%%%%%%%%%%%%%%%%%%%%%%%%%%%%%%%%%%%%%%%%%%%%%%%%%%%%%%%%%%%%%

%%%%%%%%%%%%%%%%%%%%%%%%%%%%%%%%%%%%%%%%%%%%%%%%%%%%%%%%%%%%%%%%%%%%%%%%%%%%%%%%%%%%%%%%%

We obtained 148 images for this trapezium in filters $U$--$I$, with exposure times of 0.2 and 0.5 seconds.

%%%%%%%%%%%%%%%%%%%%%%%%%%%%%%%%%%%%%%%%%%%%%%%%%%%%%%%%%%%%%%%%%%%%%%%%%%%%%%%%%%%%%%%%%

    \begin{table}[!htbp]
    \caption{Integration times for each filter for the trapezium ADS 2843}
    \begin{center}
    \begin{tabular}{ c  c  } \hline
    Filter & Time (sec) \\
    \hline
    U & 10 \\
    B & 5 \\
    V & 3 \\
    R & 2 \\
    I & 1 \\
    \hline
    \end{tabular}
    \end{center}
    \end{table}

%%%%%%%%%%%%%%%%%%%%%%%%%%%%%%%%%%%%%%%%%%%%%%%%%%%%%%%%%%%%%%%%%%%%%%%%%%%%%%%%%%%%%%%%%

The right panel  of Figure \ref{fig:ComponentsImageADS2843} shows the image of this trapezium in the $V$ filter. Component A is clearly saturated, and its brightness affects component B making it impossible to obtain good photometric measurements.

In Table \ref{tab:RadiosADS2843} we show the photometric radius for each component.

%%%%%%%%%%%%%%%%%%%%%%%%%%%%%%%%%%%%%%%%%%%%%%%%%%%%%%%%%%%%%%%%%%%%%%%%%%%%%%%%%%%%%%%%%

    \begin{table}[H]
    \caption{Photometric aperture radii in pixels for trapezium ADS 2843.}
    \begin{center}
    \begin{tabular}{ c  c  c  c } \hline
     & $r_{1}$  & $r_{2}$   & $r_{3}$ \\
     \hline & $(pix)$  & $(pix)$   & $(pix)$ \\
    \hline
    C & 8   & 10    & 12 \\
    D & 12  & 14    & 16 \\
    E & 12  & 14    & 16 \\
    \hline
    \end{tabular}
    \label{tab:RadiosADS2843}
    \end{center}
    \end{table}

%%%%%%%%%%%%%%%%%%%%%%%%%%%%%%%%%%%%%%%%%%%%%%%%%%%%%%%%%%%%%%%%%%%%%%%%%%%%%%%%%%%%%%%%%

In a similar manner as before, we use the transformation coefficients in Tables \ref{tab:Coeficientes U}--\ref{tab:Coeficientes I} and equations \ref{eq:MC0}-\ref{eq:MC4} to transform the number of counts for each star to intrinsic magnitudes and colours (see Table \ref{tab:ColyMagObsADS2843}).%De igual manera que para los trapecios anteriores, se utiliza las Ecuaciones [\ref{Ec:MC0}-\ref{Ec:MC4}] para cada filtro, tomando los coeficientes A, K y C correspondientes a la noche del 11 al 12 de diciembre. La Tabla \ref{tab:ColyMagObsADS2843} muestra los resultados de magnitud y color para ADS 2843.

%%%%%%%%%%%%%%%%%%%%%%%%%%%%%%%%%%%%%%%%%%%%%%%%%%%%%%%%%%%%%%%%%%%%%%%%%%%%%%%%%%%%%%%%%

    \begin{table}[!htbp]
    \caption{Observed magnitudes and colours for ADS 2843.}
    \begin{center}
    \begin{tabular}{ c  c  c  c  c  c  c  c  c  c  c  } \hline
     &     U &     B &     V &     R &     I &   U-B &   B-V &   V-R &   R-I &   V-I \\
    \hline
    C &  12.79 &  12.39 &  11.44 &  11.00 &  10.69 &  0.40 &  0.95 &  0.44 &  0.31 &  0.75 \\
    D &  11.49 &  11.07 &  10.22 &   9.87 &   9.67 &  0.41 &  0.85 &  0.35 &  0.20 &  0.55 \\
    E &  10.62 &  10.27 &   9.79 &   9.66 &   9.61 &  0.35 &  0.48 &  0.13 &  0.05 &  0.18 \\
    \hline
    \end{tabular}
    \label{tab:ColyMagObsADS2843}
    \end{center}
    \end{table}

%%%%%%%%%%%%%%%%%%%%%%%%%%%%%%%%%%%%%%%%%%%%%%%%%%%%%%%%%%%%%%%%%%%%%%%%%%%%%%%%%%%%%%%%%

Table \ref{tab:SpectralTypeADS2843} shows the spectral type associated to each star based on the value of its $Q$ parameter. %muestra el tipo espectral que se asocia a cada estrella de este trapecio basado en los valores calculados para el índice Q.

%%%%%%%%%%%%%%%%%%%%%%%%%%%%%%%%%%%%%%%%%%%%%%%%%%%%%%%%%%%%%%%%%%%%%%%%%%%%%%%%%%%%%%%%%

    \begin{table}[!htbp]
    \caption{$Q$ derived spectral types for ADS 2843}
    \begin{center}
    \begin{tabular}{  c  c  c  } \hline
     &         Q & Spectral Type \\ \hline
    C & -0.28 &             B8 \\
    D & -0.20 &             B9 \\
    E &  0.00 &             A0 \\
    \hline
    \end{tabular}
    \label{tab:SpectralTypeADS2843}
    \end{center}
    \end{table}

%%%%%%%%%%%%%%%%%%%%%%%%%%%%%%%%%%%%%%%%%%%%%%%%%%%%%%%%%%%%%%%%%%%%%%%%%%%%%%%%%%%%%%%%%

\begin{figure}[!htbp]
    \centering
    \includegraphics[width=10cm]{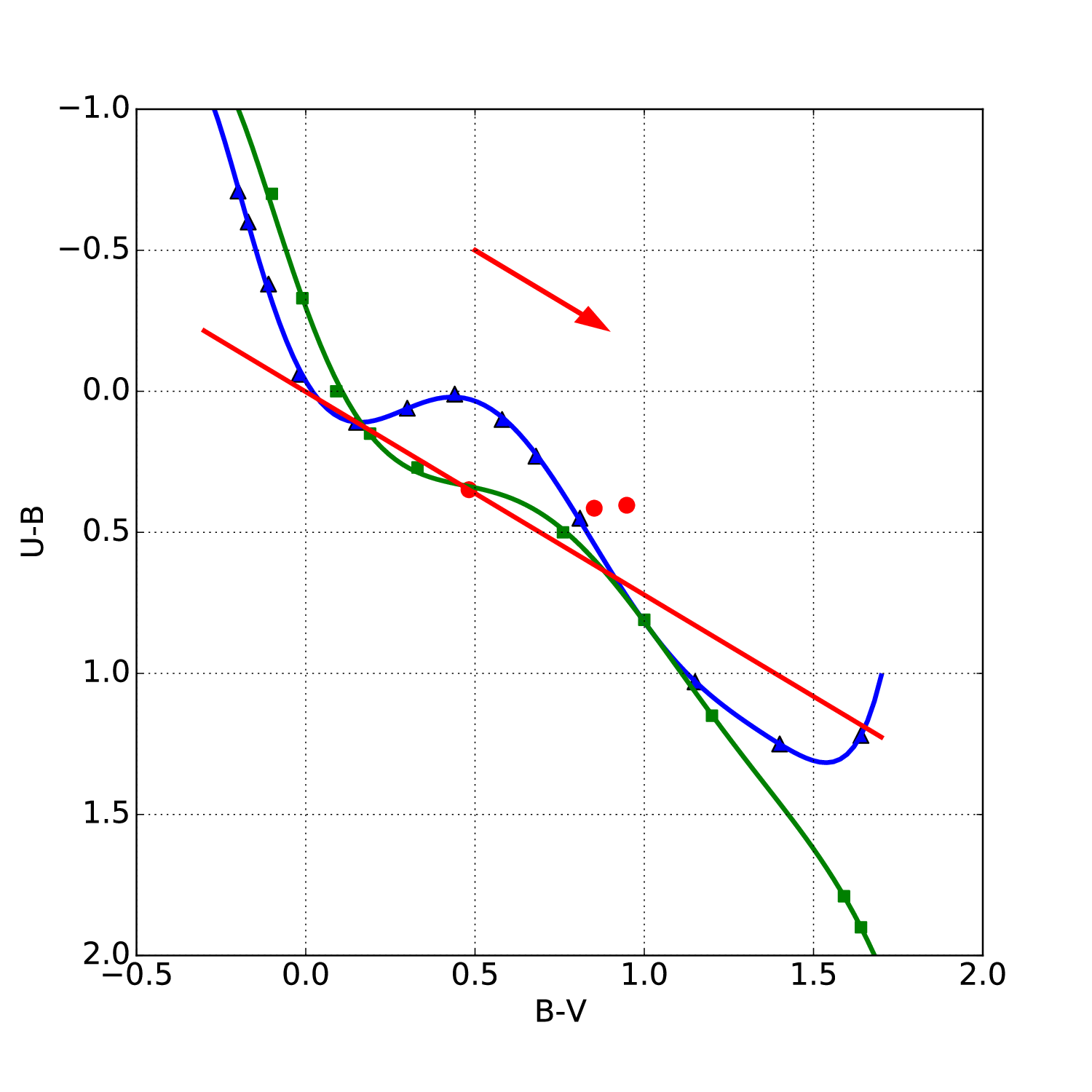}
    \caption{Two-colour diagram $(U-B\,  vs \, B-V)$ for  ADS 2843.} %Se presenta en verde la curva de estrellas supergigantes, en azul la curva de secuencia principal y en puntos rojos los colores asociados a las componentes del trapecio.}
    \label{fig:colorcolorADS2843}
   \end{figure}

Figure \ref{fig:colorcolorADS2843} shows the two-colour diagram for  ADS 2843, where the red dots represent component $C$, $D$ and $E$. %donde los puntos rojos representan las componentes C, D y E.

%%%%%%%%%%%%%%%%%%%%%%%%%%%%%%%%%%%%%%%%%%%%%%%%%%%%%%%%%%%%%%%%%%%%%%%%%%%%%%%%%%%%%%%%%

%%%%%%%%%%%%%%%%%%%%%%%%%%%%%%%%%%%%%%%%%%%%%%%%%%%%%%%%%%%%%%%%%%%%%%%%%%%%%%%%%%%%%%%%%

From the two-colour diagram we obtain the values for the intrinsic colours, excesses, absorption and distance. However, in this case, the reddening lines for the components cross the intrinsic lines more than once. We have taken the excesses that appear to be more similar. %A partir del diagrama de dos colores se construye los colores intrínsecos, excesos, absorción y distancia. Como se puede observar, en este caso las componentes cruzan más de una vez a las curvas de secuencia principal y de supergigantes, por lo que se ha tomado la intersección que haga el exceso más parecido posible.
Table \ref{tab:DistanciasADS2843} show the results for each case. %muestran los resultados para cada uno de los casos.

%%%%%%%%%%%%%%%%%%%%%%%%%%%%%%%%%%%%%%%%%%%%%%%%%%%%%%%%%%%%%%%%%%%%%%%%%%%%%%%%%%%%%%%%%

   \begin{table}[!htbp]
    \caption{Intrinsic colours, excesses, absorption and distances for ADS 2843.}
    \begin{center}
    \begin{tabular}{  c  c  c  c  c  c  c  c }
    \hline
    \multicolumn{8}{c}{Assuming Supergiant Stars}\\
    \hline
      & $(U-B)_{0}$&    $(B-V)_{0}$&    $E_{U-B}$&  $E_{B-V}$&  $A_{V}$&    Distance (pc)& Parallax (mas)\\
    \hline
    C &   -0.27 &    0.01 &    0.68 &    0.94 &  2.92 &   8989.06 &      0.11 \\
    D &   -0.17 &    0.04 &    0.58 &    0.81 &  2.51 &   6119.40 &      0.16 \\
    E &    0.12 &    0.17 &    0.22 &    0.31 &  0.96 &  10113.19 &      0.10 \\
    \hline
    \multicolumn{8}{c}{Assuming Main Sequence Stars}\\
    \hline
      & $(U-B)_{0}$&    $(B-V)_{0}$&    $E_{U-B}$&  $E_{B-V}$&  $A_{V}$&    Distance (pc)& Parallax (mas)\\
    \hline
    C &   -0.36 &   -0.11 &    0.76 &    1.06 &  3.28 &    475.21 &      2.10 \\
    D &   -0.26 &   -0.08 &    0.67 &    0.93 &  2.90 &    255.21 &      3.92 \\
    E &    0.03 &    0.04 &    0.32 &    0.45 &  1.38 &    251.98 &      3.97 \\
    \hline
    \end{tabular}
    \label{tab:DistanciasADS2843}
    \end{center}
    \end{table}

%%%%%%%%%%%%%%%%%%%%%%%%%%%%%%%%%%%%%%%%%%%%%%%%%%%%%%%%%%%%%%%%%%%%%%%%%%%%%%%%%%%%%%%%%

The magnitudes of the stars listed in SIMBAD come from the following reference: \citet{Hog2000}.
Table \ref{tab:ComparacionADS2843} shows the comparison of our results with SIMBAD (*).

%%%%%%%%%%%%%%%%%%%%%%%%%%%%%%%%%%%%%%%%%%%%%%%%%%%%%%%%%%%%%%%%%%%%%%%%%%%%%%%%%%%%%%%%%

    \begin{table}[!htbp]
    \caption{Comparison with SIMBAD (*) for ADS 2843.}
    \begin{center}
    \begin{tabular}{  c  c  c  c } \hline
    \multicolumn{4}{ c }{Magnitude} \\ \hline
     & C   & D     & E     \\ \hline
    U & 12.79   & 11.49 & 10.62 \\
    U*& -       & -     & -     \\
    B & 12.39   & 11.07 & 10.27 \\
    B*& -       & 11.05 & 10.21 \\
    V & 11.44   & 10.22 & 9.79  \\
    V*& 11.24   & 10.36 & 9.92  \\
    R & 11.00   & 9.87  & 9.66  \\
    R*& -       & 10.21 & -     \\
    I & 10.69   & 9.67  & 9.61  \\
    I*& -       & -     & -     \\ \hline
    \multicolumn{4}{ c }{Spectral Type} \\ \hline
     & C   & D     & E     \\ \hline
    St& B8      & B9    & A0   \\
    St*& -      & -     & A2V  \\ \hline
    \multicolumn{4}{ c }{Parallax} \\ \hline
     & C   & D     & E     \\ \hline
    $P_{sg}$ & 0.11  & 0.16  & 0.10 \\
    $P_{sp}$ & 2.10  & 3.92  & 3.97 \\
    P        & 3.4918 & 7.3793 & 3.5114 \\ \hline
    \end{tabular}
    \label{tab:ComparacionADS2843}
    \end{center}
    \end{table}

%%%%%%%%%%%%%%%%%%%%%%%%%%%%%%%%%%%%%%%%%%%%%%%%%%%%%%%%%%%%%%%%%%%%%%%%%%%%%%%%%%%%%%%%%

We show the magnitudes dereddened with van de Hulst curve 15 in Table \ref{tab:ColyMagDesADS2843}. This Table assumes respectively Supergiant stars and Main Sequence stars. % muestran respectivamente los valores obtenidos suponiendo estrellas supergigantes y de secuencia principal. Nótese que los valores de color coinciden con los reportados en las Tablas \ref{tab:DistanciaADS2843} y \ref{tab:DistanciaADS2843}.

%%%%%%%%%%%%%%%%%%%%%%%%%%%%%%%%%%%%%%%%%%%%%%%%%%%%%%%%%%%%%%%%%%%%%%%%%%%%%%%%%%%%%%%%%

      \begin{table}[!htbp]
     \caption{Dereddened magnitudes and colour with van de Hulst curve 15 for ADS 2843.}
    \begin{center}
    \begin{tabular}{ c c c  c  c   c  c  c  c  c  c  }
    \hline
    \multicolumn{11}{c}{Assuming Supergiant Stars}\\
    \hline
     &     U &     B &     V &     R &     I &   U-B &   B-V &   V-R &   R-I &   V-I \\
    \hline
    C &	8.27	&	8.53	&	8.52	&	8.84	&	9.3	    &	-0.26	&	0.01	&	-0.31	&	-0.46	&	-0.78\\
    D &	7.59	&	7.75	&	7.71	&	8	    &	8.47	&	-0.16	&	0.04	&	-0.29	&	-0.46	&	-0.76\\
    E &	9.12	&	8.99	&	8.82	&	8.94	&	9.15	&	0.13	&	0.17	&	-0.12	&	-0.21	&	-0.33\\
    \hline
    \multicolumn{11}{c}{Assuming Main Sequence Stars}\\
    \hline
     &     U &     B &     V &     R &     I &   U-B &   B-V &   V-R &   R-I &   V-I \\
    \hline
    C &	7.7	    &	8.04	&	8.15	&	8.56	&	9.12	&	-0.35	&	-0.11	&	-0.41	&	-0.56	&	-0.97\\
    D &	6.99	&	7.24	&	7.32	&	7.72	&	8.28	&	-0.25	&	-0.08	&	-0.39	&	-0.57	&	-0.96\\
    E &	8.48	&	8.44	&	8.41	&	8.63	&	8.95	&	0.03	&	0.04	&	-0.22	&	-0.32	&	-0.54\\
    \hline
    \end{tabular}
    \label{tab:ColyMagDesADS2843}
    \end{center}
    \end{table}
 
\vskip10cm
\vfill

%%%%%%%%%%%%%%%%%%%%%%%%%%%%%%%%%%%%%%%%%%%%%%%%%%%%%%%%%%%%%%%%%%%%%%%%%%%%%%%%%%%%%%%%%

\subsubsection{ADS 16795}
\label{subsubsec:ads16795}

\begin{figure}[!htbp]
    \centering
    \includegraphics[width=13cm]{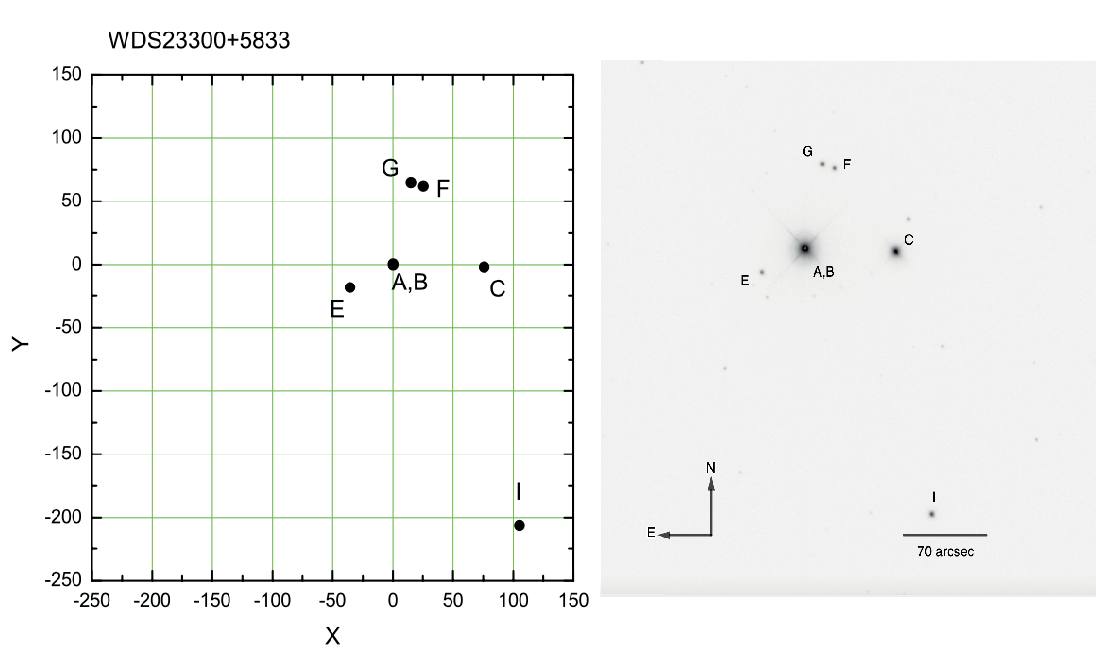}
    \caption{The left image represents the components of the trapezium ADS 16795. The axes units are in arcseconds.On the right, we show our observed image in the $V$ filter, with N-E orientation and scale in arsecs. Exposure time is $5$ seconds.}
    \label{fig:ComponentaImageADS16795}
    \end{figure}

Trapezium ADS 16795 (also known as WDS 23300+5833) is located at RA=$23h$ $30m$, Dec=$+58^o$ $32\arcmin$ and has six components: $AB$, $C$, $E$, $F$, $G$ and $I$, whose location is shown in Figure \ref{fig:ComponentaImageADS16795}.

%%%%%%%%%%%%%%%%%%%%%%%%%%%%%%%%%%%%%%%%%%%%%%%%%%%%%%%%%%%%%%%%%%%%%%%%%%%%%%%%%%%%%%%%%

%%%%%%%%%%%%%%%%%%%%%%%%%%%%%%%%%%%%%%%%%%%%%%%%%%%%%%%%%%%%%%%%%%%%%%%%%%%%%%%%%%%%%%%%%

For this trapezium we obtained 144 images in filters $U$--$I$. This trapezium was observed on the night 2019, 14-15 December. For this night the transformation coefficients $A$, $K$ and $C$ for the filter $U$ have no physical meaning (see Table \ref{tab:Coeficientes U}), so no calculation of the $U$ magnitude was possible.

Figure \ref{fig:ComponentaImageADS16795} shows the image of this trapezium in the $V$ filter.

%%%%%%%%%%%%%%%%%%%%%%%%%%%%%%%%%%%%%%%%%%%%%%%%%%%%%%%%%%%%%%%%%%%%%%%%%%%%%%%%%%%%%%%%%

It is clear that the component $AB$ is saturated so that the photometric measurements will only be performed on the other components ($C$, $E$, $F$, $G$ and $I$). 

Table \ref{tab:RadiosADS16795} shows the photometric aperture radii for each component. %muestra los radios de apertura fotométrica utilizados para cada componente.

%%%%%%%%%%%%%%%%%%%%%%%%%%%%%%%%%%%%%%%%%%%%%%%%%%%%%%%%%%%%%%%%%%%%%%%%%%%%%%%%%%%%%%%%%

    \begin{table}[!htbp]
    \caption{Photometric Aperture radii for ADS 16795.}
    \begin{center}
    \begin{tabular}{ c  c c  c } \hline
     & $r_{1}$  & $r_{2}$   & $r_{3}$ \\
     \hline     & $(pix)$  & $(pix)$   & $(pix)$ \\
    \hline
    C & 14  & 16    & 20 \\
    E & 10  & 12    & 14 \\
    F & 9   & 11    & 13 \\
    G & 9   & 11    & 13 \\
    I & 11  & 13    & 15 \\
    \hline
    \end{tabular}
    \label{tab:RadiosADS16795}
    \end{center}
    \end{table}

%%%%%%%%%%%%%%%%%%%%%%%%%%%%%%%%%%%%%%%%%%%%%%%%%%%%%%%%%%%%%%%%%%%%%%%%%%%%%%%%%%%%%%%%%

Table \ref{tab:ColyMagObsADS16795} gives the observed magnitudes and colours for the components of this trapezium. We obtained only those for the filters $B$, $V$, $R$ and $I$ because the results for $U$ have no physical meaning and have been omitted. %muestra los resultados de magnitudes y colores observados para los filtros B, V, R e I. Se han utilizado los coeficientes A, K y C de la noche del 14 al 15 de diciembre, donde la Tabla \ref{tab:Coeficientes U} muestra que los resultados para U surgen de una mala estadística por lo que se ha omitido este filtro.

%%%%%%%%%%%%%%%%%%%%%%%%%%%%%%%%%%%%%%%%%%%%%%%%%%%%%%%%%%%%%%%%%%%%%%%%%%%%%%%%%%%%%%%%%

    \begin{table}[!htbp]
     \caption{Observed magnitudes and colours for ADS 16795.}
    \begin{center}
    \begin{tabular}{ c  c  c  c  c  c  c  c  c } \hline
     &     B &     V &     R &     I &   B-V &   V-R &   R-I &   V-I \\
    \hline
    C &   7.81 &   7.79 &   8.45 &   8.26 &  0.01 &  0.18 & -0.65 & -0.47 \\
    E &  11.83 &  11.18 &  10.74 &  10.34 &  0.65 &  0.40 &  0.44 &  0.84 \\
    F &  11.52 &  10.98 &  10.59 &  10.32 &  0.54 &  0.27 &  0.38 &  0.65 \\
    G &  11.64 &  11.08 &  10.68 &  10.40 &  0.56 &  0.28 &  0.40 &  0.67 \\
    I &  10.19 &   9.74 &   9.41 &   9.20 &  0.45 &  0.22 &  0.33 &  0.55 \\
    \hline
    \end{tabular}
    \label{tab:ColyMagObsADS16795}
    \end{center}
    \end{table}

%%%%%%%%%%%%%%%%%%%%%%%%%%%%%%%%%%%%%%%%%%%%%%%%%%%%%%%%%%%%%%%%%%%%%%%%%%%%%%%%%%%%%%%%%

Since we cannot obtain a value of $U$, it is impossible to calculate the value of the $Q$ parameter. 

The magnitudes of the stars listed in SIMBAD come from the following reference: \citet{Zacharias2012}.

In Table \ref{tab:ComparacionADS16795} we compare our results with those listed in SIMBAD.

%%%%%%%%%%%%%%%%%%%%%%%%%%%%%%%%%%%%%%%%%%%%%%%%%%%%%%%%%%%%%%%%%%%%%%%%%%%%%%%%%%%%%%%%%

    \begin{table}[!htbp]
    \caption{Comparison of our results for ADS 16795 and those listed in SIMBAD (*).}
    \begin{center}
    \begin{tabular}{  c  c  c  c  c  c } \hline
    \multicolumn{6}{ c }{Magnitude} \\ \hline
     & C    & E     & F     & G     & I \\ \hline
    B & 7.81    & 11.83 & 11.52 & 11.64 & 10.19 \\
    B*& -       & 11.93 & -     & -     & -     \\
    V & 7.79    & 11.18 & 10.98 & 11.08 & 9.74  \\
    V*& -       & 11.19 & -     & -     & -     \\
    R & 8.45    & 10.74 & 10.59 & 10.68 & 9.41  \\
    R*& 8.15    & 11.14 & 10.94 & 11.03 & -     \\
    I & 8.26    & 10.34 & 10.32 & 10.40 & 9.20  \\
    I*& -       & -     & -     & -     & -     \\ \hline
    \end{tabular}
  \label{tab:ComparacionADS16795}
    \end{center}
    \end{table}

%%%%%%%%%%%%%%%%%%%%%%%%%%%%%%%%%%%%%%%%%%%%%%%%%%%%%%%%%%%%%%%%%%%%%%%%%%%%%%%%%%%%%%%%%

\vskip10cm

%%%%%%%%%%%%%%%%%%%%%%%%%%%%%%%%%%%%%%%%%%%%%%%%%%%%%%%%%%%%%%%%%%%%%%%%%%%%%%%%%%%%%%%%%

\section{The slope of the reddening line}
\label{sec:slope}

Since the $Q$-derived spectral types differ slightly from those listed in SIMBAD, we speculate that these differences might be due to slight differences in the value of the slope of the reddening line on the two-colour diagram from the canonical value $(0.72)$.

The slope of the reddening line corresponds to the ratio of the colour excesses $E(U-B)$ and $E(B-V)$, and its value is clearly dependent on the physical and chemical properties of the interstellar medium through which the light from the stars travels on its way to our observing instruments. Carrying out a full investigation as to whether the slope of the reddening line differs from region to region is not only a rather interesting endeavour, but one that is clearly outside the scope of this paper. Here we shall only speculate that this might be the reason for the slight differences in stellar spectral types and use the seven points we have to obtain a value for the new slope. \citet{Aidelman2023} suggest that the possible difference of the slope of the reddening line from the canonical value, may be due to an anomalous colour excess or variations of the extinction law.   

%Those authors also suggests that the possible differences in the value of the slope of the
%reddening line may occur if the star has an anomalous color excess. So these Q
%parameters may help to detect the peculiarity of the extinction law towards particular sky
%regions.

In Table \ref{tab:QobsVSQint} we present the object name, the Q-derived Spectral Type, the measured $Q$ index, the SIMBAD Spectral Type and the value of the $Q$ index for the Spectral Type given in column 4.

%%%%%%%%%%%%%%%%%%%%%%%%%%%%%%%%%%%%%%%%%%%%%%%%%%%%%%%%%%%%%%%%%%%%%%%%%%%%%%%%%%%%%%%%%

    \begin{table}[!htbp]
     \caption{$Q$ parameter values associated with the spectral value derived and the types given in the literature.}
    \centering
    \begin{tabular}{ccccc}\hline
$Object$	    &	$ST_{obs}$	&	$Q_{obs}$	&	$ST_{int}$	&	$Q_{int}$	\\ \hline
%ADS15184AB	&	B5	&	-0.50	&	-	&	-	    \\
ADS15184C	&	B5	&	-0.48	&	B1.5&	-0.74	\\
ADS15184D	&	B6	&	-0.40	&	B1	&	-0.78	\\
ADS4728C	&	B8	&	-0.20	&	B9	&	-0.13	\\
ADS4728D	&	B3	&	-0.63	&	B1	&	-0.78	\\
ADS4728E	&	B5	&	-0.42	&	B3	&	-0.57	\\
ADS4728F	&	B9	&	-0.08	&	B9	&	-0.13	\\
%ADS4728G	&	A0	&	-0.03	&	-	&	-	    \\
%ADS4728H	&	A0	&	0.30	&	G8	&	-	    \\
ADS4728I	&	B8	&	-0.26	&	B5	&	-0.44	\\
%ADS4728J	&	B7	&	-0.32	&	-	&	-	    \\
%ADS2843C	&	B8	&	-0.28	&	-	&	-	    \\
%ADS2843D	&	B9	&	-0.20	&	-	&	-	    \\
%ADS2843E	&	A0	&	0.00	&	A2	&	-	    \\
%ADS13292A	&	B3	&	-0.6	&	B1	&	-0.78	\\
%ADS13292B	&	B5	&	-0.44	&	OB	&	-	    \\
%ADS13292C	&	B5	&	-0.46	&	-	&	-	    \\
%ADS13292D	&	B5	&	-0.47	&	-	&	-	    \\
%ADS13292E	&	B8	&	-0.20	&	-	&	-	    \\
%ADS13292F	&	B9	&	-0.07	&	-	&	-	    \\
%ADS13374A	&	B2	&	-0.68	&	WN	&	-	    \\
%ADS13374D	&	B3	&	-0.61	&	B1	&	-0.78	\\
%ADS13374E	&	B5	&	-0.43	&	-	&	-	    \\
%ADS13374F	&	B2	&	-0.67	&	B0	&	-0.90	\\
    \hline
    \end{tabular}
    \label{tab:QobsVSQint}
    \end{table}

%%%%%%%%%%%%%%%%%%%%%%%%%%%%%%%%%%%%%%%%%%%%%%%%%%%%%%%%%%%%%%%%%%%%%%%%%%%%%%%%%%%%%%%%%

Figure \ref{fig:QobsVSQint} shows the relation we obtain from Table \ref{tab:QobsVSQint} between the observed $Q$ index ($Q_{obs}$) and the $Q$ associated to the Spectral Type reported in SIMBAD ($Q_{int}$). We fitted a least squares straight line that produced the following equation:

%%%%%%%%%%%%%%%%%%%%%%%%%%%%%%%%%%%%%%%%%%%%%%%%%%%%%%%%%%%%%%%%%%%%%%%%%%%%%%%%%%%%%%%%%

   \begin{figure}[H]
    \centering
    \includegraphics[width=10cm]{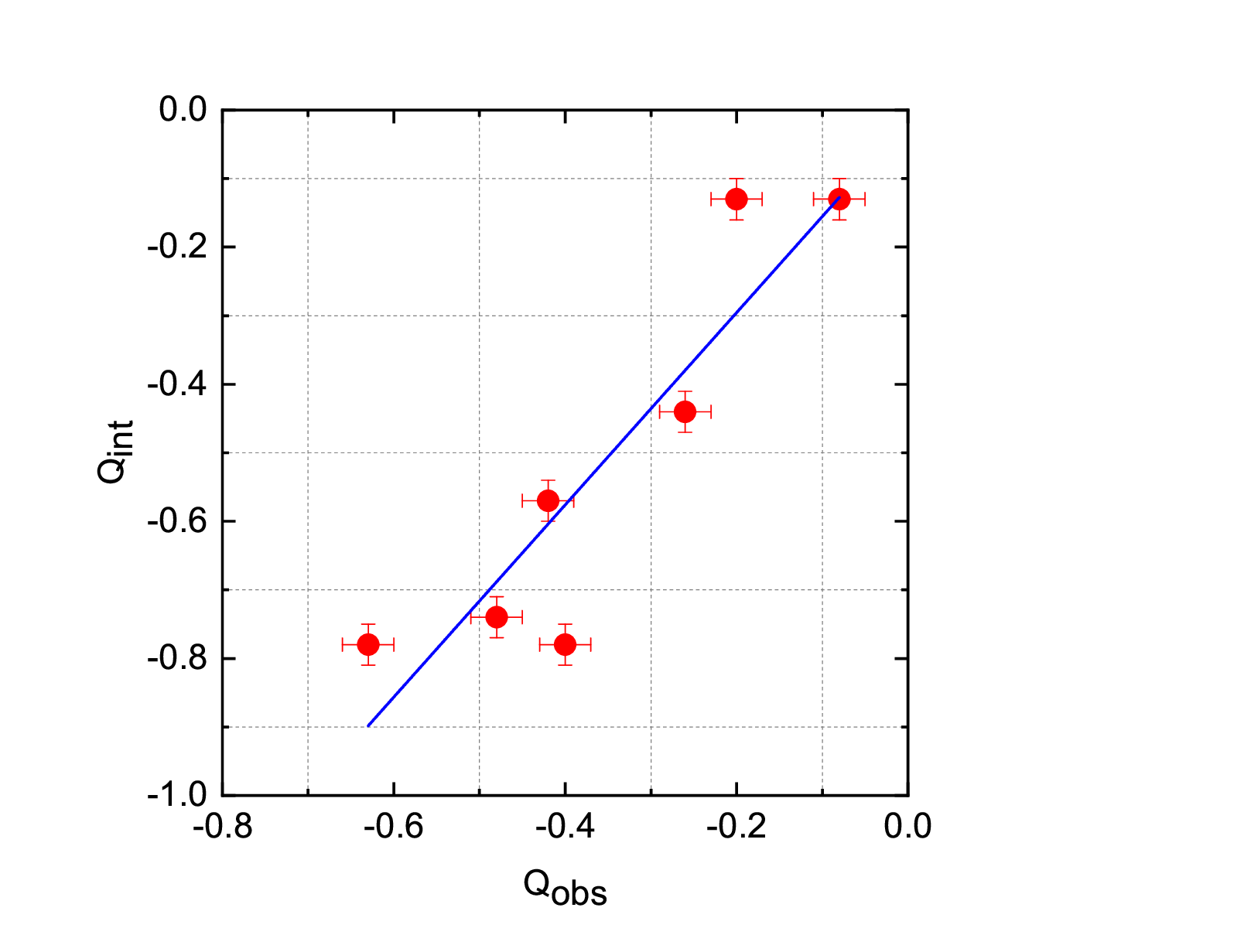}
    \caption{Graph of the values shown in Table \ref{tab:QobsVSQint}.}
    \label{fig:QobsVSQint}
    \end{figure}

%%%%%%%%%%%%%%%%%%%%%%%%%%%%%%%%%%%%%%%%%%%%%%%%%%%%%%%%%%%%%%%%%%%%%%%%%%%%%%%%%%%%%%%%%

    \begin{equation}
    Q_{int}=(1.40 \pm 0.30) \ Q_{obs} - (0.02 \pm 0.12).
    \label{eq:QobsVSQint}
    \end{equation}

%%%%%%%%%%%%%%%%%%%%%%%%%%%%%%%%%%%%%%%%%%%%%%%%%%%%%%%%%%%%%%%%%%%%%%%%%%%%%%%%%%%%%%%%%

    Using this equation we might be able to correct the $Q$-derived Spectral Types, however, the number of points which produce this equation is very small so, at present, we have decided to leave its use and confirmation of usefulness for further investigations.
    
    Algebraic manipulation of equation (\ref{eq:QobsVSQint}) allows us to express it as follows:

    \begin{equation}
        1.40E(B-V)\ m+0.40\ (U-B)_{int}-1.01\ (B-V)_{Obs}+0.72\ (B-V)_{int}-0.02=0
        \label{eq:difslope}
    \end{equation}

    where $m$ represents the possible new slope of the reddening line, and the suffixes $int$ and $Obs$ refer to the intrinsic and observed values. Effecting a least squares solution for $m$ from this equation would allow us to explain the difference between the Q-derived Spectral types and the Spectral types listed in SIMBAD as a difference (from 0.72) of the slope of the reddening line on the two-colour diagram.
    
   In what follows we shall use the information we have for the seven discrepant points and calculate a value for the slope of the reddening line that fits these results. In Table \ref{tab:slopereddening} we see in Column 1 the name of the star, in Column 2 the Spectral type, in Column 3  the $(U-B)_{int}$, in Column 4 $(B-V)_{Obs}$, in Column 5 $(B-V)_{int}$ and in Column 6  $E(B-V)$ for the stars for which a discrepant Q-derived spectral type was found. 
    
    %$E(B-V)$ We expect to collect more data in order to be able to produce a more significant solution to equation (\ref{eq:difslope}).

%%%%%%%%%%%%%%%%%%%%%%%%%%%%%%%%%%%%%%%%%%%%%%%%%%%%%%%%%%%%%%%%%%%%%%%%%%%%%%%%%%%%%%%%%

   \begin{table}[!htbp]
     \caption{Data for stars with discrepant Spectral types between the Q-derived value and the value found in SIMBAD.}
    \centering
    \begin{tabular}{cccccc}\hline
$Object$	    &	$Spectral\  Type$	&	$(U-B)_{int}$	&	$(B-V)_{obs}$	&	$(B-V)_{int}$  &  $E(B-V)$	\\ \hline

ADS15184C	&	B1.5V	            &	-0.88           &	0.32            &	-0.25          &     0.57   \\
ADS15184D	&	B61V                &	-0.95           &	0.25            &	-0.26	       &     0.51   \\
ADS4728C	&	B9V	                &	-0.18	        &   0.21	        &	-0.07	       &     0.28   \\
ADS4728D	&	B1V	                &	-0.95	        &	0.01	        &	-0.26	       &     0.27   \\
ADS4728E	&	B3V	                &	-0.68	        &	0.10	        &	-0.20	       &     0.30    \\
ADS4728F	&	B9V	                &	-0.18	        &	0.21	        &	-0.07	       &     0.28     \\
ADS4728I	&	B8	                &	-0.36	        &	0.12	        &	-0.11	       &     0.23     \\

    \hline
    \end{tabular}
    \label{tab:slopereddening}
    \end{table}

%%%%%%%%%%%%%%%%%%%%%%%%%%%%%%%%%%%%%%%%%%%%%%%%%%%%%%%%%%%%%%%%%%%%%%%%%%%%%%%%%%%%%%%%%

    Effecting a least squares solution for $m$ in equation (\ref{eq:difslope}) using the values listed in Table (\ref{tab:slopereddening}) produces a value for the slope of the reddening line $m=1.14^{+0.34}_{-0.61}$ higher than the accepted one. However, the number of points which we are using for the calculation is rather limited and we could hardly expect the result to have some degree of statistical significance. We would like to leave this fact as a possible explanation for the discrepancy of spectral types and hope that, as more discrepant results are collected with the observations of further trapezia stars, the determination of this new slope for the reddening line acquires a more robust statistical significance.

%%%%%%%%%%%%%%%%%%%%%%%%%%%%%%%%%%%%%%%%%%%%%%%%%%%%%%%%%%%%%%%%%%%%%%%%%%%%%%%%%%%%%%%%%
    
\section{Conclusions}
\label{sec:conclusions}

This paper constitutes a photometric study of some trapezia in the Galaxy. The data were obtained during several observing seasons. 

Here we present CCD photometry of the brighter stars in the stellar trapezia $ADS\ 15184$, $ADS \ 4728$, $ADS \ 2843$, and $ADS \ 16795$, which we have used to explore the possibility of finding the spectral type of the stars using the $Q$ parameter, defined as $Q=(U-B)-0.72(B-V)$. This parameter is reddening independent, since the slope of the reddening line on the two-colour diagram is approximately equal to $0.72$  (see Johnson \& Morgan (1953)). This, of course, is a first approach since the calibration we used for the $Q$ parameter versus the Spectral Type does not take into consideration peculiar stars and is only based on typical stars of luminosity classes $I$ and $V$.

The spectral types which we have determined coincide reasonably well with those listed in SIMBAD. However, a different value of the slope of the reddening line on the two-colour diagram might produce a better coincidence between the $Q$-derived spectral types and those listed in SIMBAD. Effecting a least squares solution for the slope of the reddening line in equation (\ref{eq:difslope}) produces a value of its slope $m$ of $1.14^{+0.34}_{-0.61}$, ascribing the difference in spectral types to a difference in the slope of the reddening line is, at this point, only a speculation which needs a full investigation to be confirmed or discarded. The $Q$ parameter appears to be useful only for early type stars (earlier than A0), so if it is used for stars whose spectral type is later than A0 the results obtained might be in error.

As part of our study we intend to determine the spectral type of the stars in the trapezia through classification of their spectra and hope to be able to measure their radial velocities which, joined with proper motions from GAIA will allow us to perform a detailed dynamical study of the Galactic trapezia.

%%%%%%%%%%%%%%%%%%%%%%%%%%%%%%%%%%%%%%%%%%%%%%%%%%%%%%%%%%%%%%%%%%%%%%%%%%%%%%%%%%%%%%%%%

%%%%%%%%%%%%%%%%%%%%%%%%%%%%%%%%%%%%%%%%%%%%%%%%%%%%%%%%%%%%%%%%%%%%%%%%%%%%%%%%%%%%%%%%%

\section{Acknowledgements}

We would like to thank the Instituto de Astronom\'ia at Universidad Nacional Aut\'onoma de M\'exico (IAUNAM) and the Instituto de Astronom\'ia y Metereolog\'ia at Universidad de Guadalajara (IAMUdeG) for providing a congenial and stimulating atmosphere in which to work. We also thank the computing staff at both institutions for being always available and ready to help with random problems with our computing equipment, which arise when one least expects them. We would also like to thank Juan Carlos Yustis for help with the production of the figures in this paper.  We also thank Direcci\'on General de Asuntos del Personal Acad\'emico, DGAPA at UNAM for financial support under projects PAPIIT IN103813, IN102517 and IN102617. This research has made use of the SIMBAD database, operated at CDS, Strasbourg, France. The help and valuable suggestions provided by an anonymous referee are gratefully acknowledged.

%%%%%%%%%%%%%%%%%%%%%%%%%%%%%%%%%%%%%%%%%%%%%%%%%%%%%%%%%%%%%%%%%%%%%%%%%%%%%%%%%%%%%%%%%

 \bibliography{REFERENCIAS}

%%%%%%%%%%%%%%%%%%%%%%%%%%%%%%%%%%%%%%%%%%%%%%%%%%%%%%%%%%%%%%%%%%%%%%%%%%%%%%%%%%%%%%%%%

%%%%%%%%%%%%%%%%%%%%%%%%%%%%%%%%%%%%%%%%%%%%%%%%%%%%%%%%%%%%%%%%%%%%%%%%%%%%%%%%%%%%%%%%%

 \end{document}